# Multi-resonant non-dispersive infrared gas sensing: breaking the selectivity and sensitivity tradeoff

Emma R. Bartelsen[1,2], J. Ryan Nolen[3], Christopher R. Gubbin[3], Mingze He[2,4], Ryan W. Spangler[5], Joshua Nordlander[5], Katja Diaz-Granados[1], Simone De Liberato[3,6,7], Jon-Paul Maria[5], James R. McBride[8], Joshua D. Caldwell[1,2,3]

[1] Interdisciplinary Materials Science Program, Vanderbilt University, Nashville, Tennessee 37240, USA
[2] Department of Mechanical Engineering, Vanderbilt University, Nashville, Tennessee 37235, USA
[3] Sensorium Technological Labs, 6714 Duquaine Ct, Nashville, Tennessee 37205, USA
[4] Photonics Initiative, Advanced Science Research Center, City University of New York, New York, NY 10031, USA
[5] Department of Materials Science and Engineering, The Pennsylvania State University, University Park, Pennsylvania 16802, USA
[6] Istituto di Fotonica e Nanotecnologie, Consiglio Nazionale delle Ricerche (CNR), Piazza Leonardo da Vinci 32, Milano, 20133, Italy
[7] School of Physics and Astronomy, University of Southampton, University Road, Southampton, SO17 1BJ, United Kingdom
[8] Department of Chemistry, The Vanderbilt Institute of Nanoscale Science and Engineering, Vanderbilt University, Nashville, TN 37235, USA

**Abstract.** In applications such as atmospheric monitoring of greenhouse gases and pollutants, the detection and identification of trace concentrations of harmful gases is commonly achieved using non-dispersive infrared (NDIR) sensors. These devices typically employ a broadband infrared emitter, thermopile detector, and a spectrally selective bandpass filter tuned to the vibrational resonance of the target analyte. However, the fabrication of these filters is costly and limited to a single frequency. This limitation introduces a fundamental tradeoff, as broadening the optical passband width enhances sensitivity but compromises selectivity, whereas narrowing improves selectivity at the expense of sensitivity. In this work, we validate a filterless NDIR gas sensing approach utilizing a multi-peak thermal emitter developed through inverse design. This emitter enhances detection sensitivity by simultaneously targeting multiple absorption bands, demonstrated through the creation of a sensor designed for the C-H vibrational modes of propane ($C_3H_8$). Additionally, a second set of single-peak emitters were developed to showcase the capability of designing highly selective sensors operating within close spectral proximity. These emitters, targeting the stretching modes of carbon monoxide (CO) and carbon dioxide ($CO_2$), exhibit quality factors (Q-factors) above 50 and minimal crosstalk, enabling accurate detection of the target gas without interference from gases with spectrally adjacent absorption bands. This is enabled by the implementation of an aperiodic distributed Bragg reflectors (a-DBRs), which allows for higher Q-factors with fewer layers than a periodic Bragg reflector using the same materials and number of layers, thereby reducing fabrication complexity and cost. Experimental results validate that this approach breaks the tradeoff between sensitivity and selectivity. This work highlights the potential of optimized thermal emitters for more efficient and compact gas sensing applications.

**Introduction**. The characteristic, infrared-active, vibrational resonances of a molecule populate the mid-infrared (MIR) spectral range, forming a unique spectral "fingerprint". This "fingerprint" provides insights about the molecular structure and composition.[1] Optical techniques are commonly employed to exploit this MIR "fingerprint" to determine the presence and concentration of gas-phase chemical species in industrial[2], medical[3–5], defense[6,7], and research settings.[8] Some applications include atmospheric sensing of greenhouse gases[9,10] and other pollutants[11] and monitoring of manufacturing gases[12] and leak detection for workplace safety. While spectroscopic techniques such as Fourier transform infrared (FTIR) spectroscopy are widely used in laboratory settings and some industrial environments, their large footprint and high cost, together with the interferometer's sensitivity to vibrations, can limit practicality in compact, field deployable sensing systems. Furthermore, these techniques often require post-processing to extract specific information such as gas classification and concentration, making them less suitable for real-time, application-specific sensing.[13,14]

To address the size and cost limitations of traditional spectroscopic techniques, many commercial applications employ non-dispersive infrared (NDIR) gas sensors.[2] These are compact optical devices that exploit the Beer-Lambert law[15,16] **[Eqn. 1]:**

$$A = \varepsilon bC, \tag{1}$$

where $A$ is the absorbance, $\varepsilon$ is the molar absorptivity, $b$ is the path length, and $C$ is the concentration. Thus, the concentration of a gas can be determined via a reduction in transmitted light, where absorption at resonant vibrational frequencies over a defined path length and cross-sectional area is directly proportional to the concentration of that molecule.

Traditional NDIR systems typically incorporate a broadband infrared light source, a thermopile or pyroelectric detector, and a spectrally selective bandpass filter tuned to one of the strongest vibrational resonances for the target gas, within a gas cell of a defined optical pathlength. One of the main limitations of NDIR systems is that these filters are limited to a single spectral band. While a filter with a sufficiently narrow passband enables single-gas selectivity, it compromises overall sensitivity by blocking a large fraction of the thermally emitted radiation, often including some of the absorption band of the molecule of interest. Systems capable of detecting multiple gas types often incorporate a rotating filter wheel with multiple bandpass filters, which increases size, complexity, and cost of the sensor. A potential solution to mitigate the limitations of single frequency bandpass filters is to integrate narrowband detectors into the system. Such narrowband detectors have been demonstrated using materials such as graphene resonators or plasmonic structures.[17,18] However, they often require nanoscale lithographic fabrication, which increases cost and limits scalability. Another alternative is the use of narrowband sources, such as quantum cascade lasers (QCLs) that offer high performance but come with high power demands and are not cost-effective for widespread industrial use.[19] Mid-infrared LEDs have gained interest due to their compact form, but they suffer from low output power and broadband emission profiles, which may potentially result in spectral crosstalk between different gas types, resulting in false positives.[20,21] Recent advancements in nanophotonics have enabled the development of narrowband infrared sources using photonic crystals[22–24], structured polaritonic materials[25–30], and metamaterial-based designs[31–33]. These nanophotonic infrared emitting metamaterials (NIREMs) can exhibit linewidths approaching those of molecular vibrational features, enabling more sensitive and selective gas detection.[27,30,34,35] However, these systems still rely on complex fabrication processes such as electron beam lithography to pattern polaritonic nanostructures and stimulate polariton resonances in the MIR.

Tamm plasmon polaritons (TPPs) are formed at the interface between a distributed Bragg reflector (DBR) and a plasmonic medium, supporting omni-directional, spectrally narrow resonances that can be engineered for selective MIR emission. Early experiments by Yang et al. showed narrowband, wavelength-selective thermal emission from Tamm structures, establishing their potential as compact, lithography-free emitters for sensing.[36] Building on these findings, our recent work demonstrated deterministic inverse design of

Tamm-based thermal emitters with single and multi-resonant control using aperiodic Bragg stacks on tunable CdO, making them well suited for gas sensing applications.[37] We further showed that coupling Tamm plasmon and Tamm phonon polaritons yields high-Q, multiband planar absorbers with fewer Bragg layers, simplifying fabrication while preserving narrow linewidths.[35] Together, these studies indicated the promise of enhanced, filterless NDIR sensing, but this had not yet been validated, motivating the approach detailed below.

In this work, we present a lithography-free, multi-frequency NDIR approach that relies on planar frequency-selective emitters supported by dielectric stacks. Using the stochastic gradient descent inverse design algorithm referenced above[37], we demonstrate control over thermal emission profiles tailored to spectral targets of varying complexity. These filterless NDIR schemes improve efficiency by eliminating off-resonant emission losses and reduce design complexity, supporting compact integration and the potential for multi-gas sensing on a single platform. Our design overcomes the limitations of conventional NDIR systems by eliminating the need for bandpass filters, narrowband detectors, or lithographically patterned metasurfaces, while preserving a compact form factor, cost efficiency, planar emitter design, and the sensitivity required for real-world gas sensing applications.

**Results.**

The emission wavelength, amplitude, and linewidth of periodic DBR-based emitters are determined by four parameters: the thickness of the constituent layers within a period, the number of periods, the materials chosen for the high- and low-refractive index layers, and the optical properties of the adjacent reflective film.[38] Achieving spectrally narrow emission with periodic stacks often requires a large number of layers to strengthen confinement and narrow the linewidth. This, in turn, complicates fabrication by demanding high precision in each layer, increasing cost, and raising the probability of interfacial defects from the many material interfaces. Furthermore, periodic stacks cannot generate arbitrary multi-peak emission, only a single resonance and its harmonics. To achieve independent control over multiple peaks, the periodicity must be broken. Breaking the periodic structure allows each layer thickness to be tuned independently, yielding aperiodic DBRs (a-DBRs) that provide far greater design flexibility and can realize complex spectral responses with fewer layers than periodic stacks.[35,37] However, the expanded design space introduces substantial complexity, making manual forward, intuition based design approaches infeasible and necessitating the use of inverse design or machine learning techniques to efficiently identify structures that meet desired spectral targets.

To address the expanded parameter space, we utilized our group's stochastic gradient descent inverse design algorithm to realize emitters with tailored, narrowband thermal emission profiles.[37] This approach provides precise control over emission wavelengths and linewidths, enabling simultaneous targeting of multiple molecular vibrational modes. Such multi-resonant profiles improve sensitivity without compromising selectivity in a filterless NDIR configuration. To validate this approach, we performed two experiments. First, we designed a dual-peak emitter targeting the C-H deformation (1300 - 1500 cm$^{-1}$) and stretching (2800 - 3000 cm$^{-1}$) modes of propane ($C_3H_8$)[39], enabling filterless NDIR sensing of both resonances in a single measurement, as illustrated in **Fig. 1(b)**. Second, we designed single-peak emitters for carbon monoxide (CO) and carbon dioxide ($CO_2$), targeting the stretching mode at 2150 cm$^{-1}$ and the asymmetric stretching mode at 2349 cm$^{-1}$, respectively, to test spectral selectivity for closely spaced absorption features.[40,41] Both emitters exhibited Q-factors above 50, effectively isolating their target resonances and minimizing spectral crosstalk. Combined, we quantify and validate the enhancement in NDIR sensitivity through multi-mode targeting, while maintaining high selectivity, as confirmed by the lack of detection of non-target gases with absorption bands in close spectral proximity.

**$C_3H_8$ Emitter.** To enable multi-resonant emission for $C_3H_8$ detection, we fabricated a thermal emitter comprising an a-DBR stack of eight alternating Ge and $AlO_x$ layers on a low-resistivity Si substrate (layer design shown in **[Fig. 1(a)]**). Target and as-grown layer thicknesses are provided in the Supporting Information. The a-DBR was designed to produce thermal emission at two distinct peaks, with the lower-frequency band centered at 1420 cm$^{-1}$ (C-H deformation) and a higher-frequency band centered at 2768 cm$^{-1}$ (C-H stretching) **[Fig. 1(c)]**. These peaks align closely with the target and calculated spectra **[Fig. 1(d)]**, demonstrating the strong spectral tunability of this approach and its potential for engineering emitters with precisely tailored spectral profiles.

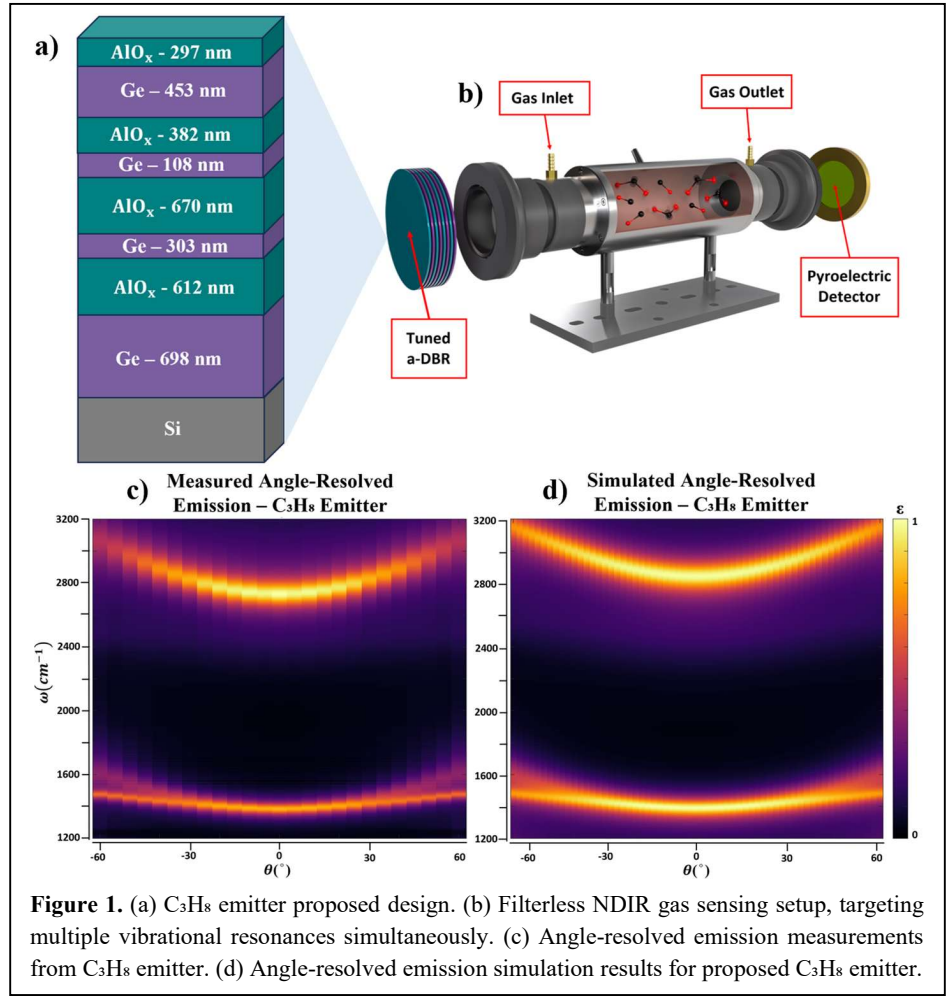

**Figure 1.** (a) $C_3H_8$ emitter proposed design. (b) Filterless NDIR gas sensing setup, targeting multiple vibrational resonances simultaneously. (c) Angle-resolved emission measurements from $C_3H_8$ emitter. (d) Angle-resolved emission simulation results for proposed $C_3H_8$ emitter.

Along the surface normal, the emitter exhibited strong overlap with the C-H deformation mode of $C_3H_8$, with emissivity $\varepsilon \approx 0.98$ at 1420 cm$^{-1}$ (FWHM = 61.15 cm$^{-1}$). The higher-frequency band peaked at 2768 cm$^{-1}$ ($\varepsilon \approx 0.98$, FWHM = 29.5 cm$^{-1}$), slightly red-shifted from the C-H stretching mode. Both peaks blue-shifted with angle, with the lower-frequency band showing a small curvature ($b_l$ = 0.0278 $\frac{cm^{-1}}{deg^2}$) that maintained strong overlap with the $C_3H_8$ absorption feature up to 60°. The higher-frequency band exhibited a larger curvature ($b_u$ = 0.2583 $\frac{cm^{-1}}{deg^2}$), maintaining overlap up to 33° **[Fig. 1(c)]**. These angle-resolved measurements were acquired using a custom-built FTIR-based heated rotation stage, described in greater detail in the Methods section. The measured angle-resolved emissivity spectra at 300 °C showed excellent agreement with transfer-matrix-based absorptivity calculations **[Fig. 1(d)]**.[42] Details regarding the thermal emission measurements, including the schematic of the experimental setup, can be found in the Supporting Information.

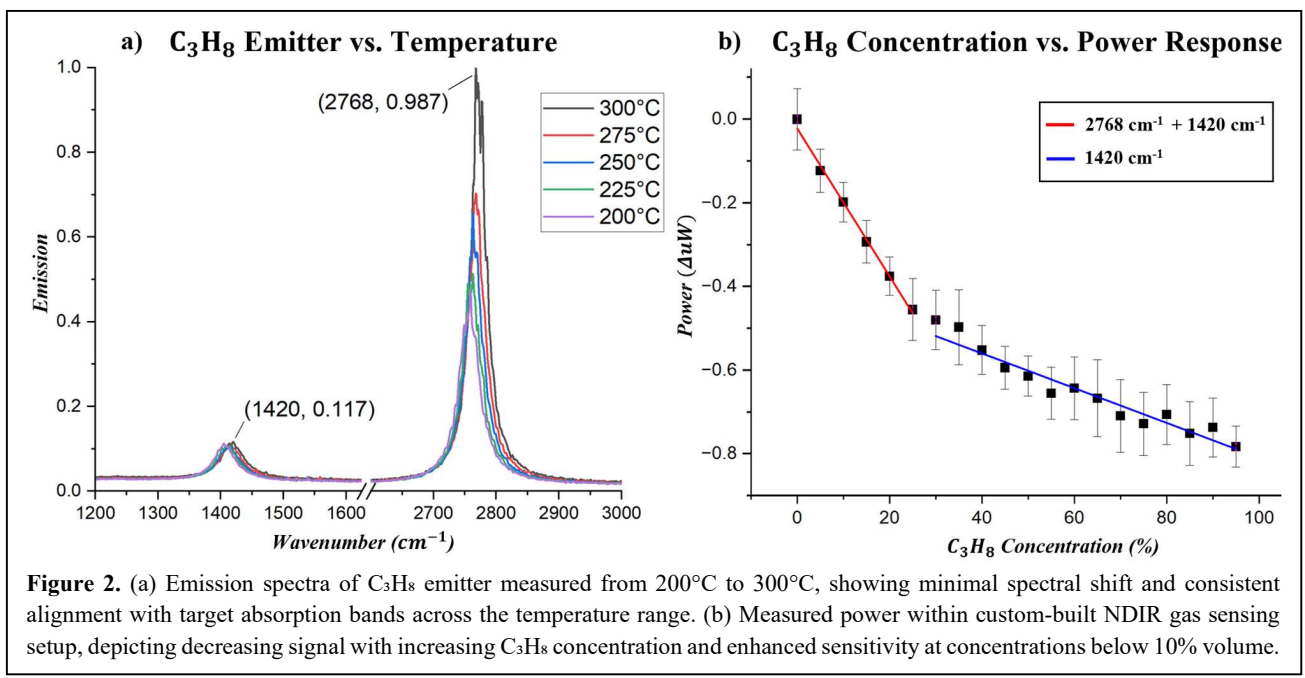

**Figure 2.** (a) Emission spectra of $C_3H_8$ emitter measured from 200°C to 300°C, showing minimal spectral shift and consistent alignment with target absorption bands across the temperature range. (b) Measured power within custom-built NDIR gas sensing setup, depicting decreasing signal with increasing $C_3H_8$ concentration and enhanced sensitivity at concentrations below 10% volume.

To assess the thermal stability of the emitter, we measured its spectral output from 200°C to 300°C using our FTIR system **[Fig. 2(a)]**. The emission spectra show minimal spectral shifting across this temperature range, with peak positions remaining well-aligned with the vibrational absorption bands of $C_3H_8$. This consistency is critical for reliable sensing, as it ensures the emitted wavelengths remain within the absorption features of the target analyte.

We then evaluated the dual-peak emitter in a custom NDIR gas sensing setup (schematic in Supporting Information). The emitter was operated at 300 °C, and off-normal emission was collected with a Winston cone to maximize overlap with the $C_3H_8$ absorption features. Propane concentrations were varied from 0 to 400,000 ppm (0 - 40% by volume). As the concentration increased, the detected power decreased in accordance with Beer-Lambert behavior **[Fig. 2(b)]**. The dual-band design enhanced sensitivity in the 0 - 100,000 ppm range (0 - 10%), yielding a slope approximately 4.25× steeper than at higher concentrations. At concentrations above 10%, the 2768 cm$^{-1}$ band saturated, leaving the response dominated by the lower-frequency mode. These results highlight the advantage of the dual-band design, which improves sensitivity in the 0 - 10% range compared to conventional single-band NDIR systems while maintaining a wide dynamic range. Simulations confirm strong overlap between the emitter output and the vibrational resonances of $C_3H_8$ at 1420 cm$^{-1}$ and 2768 cm$^{-1}$, with both bands contributing to absorbance and producing a larger power drop than either band alone (Supporting Information).

To evaluate performance relative to commercial systems, the propane emitter was compared to a single-band InfraTec NDIR filter centered at 2678 cm$^{-1}$ with a FWHM of 63.5 cm$^{-1}$.[43] The filter's broader 63.5 cm$^{-1}$ FWHM, compared to our emitter's 29.5 cm$^{-1}$ linewidth at 2768 cm$^{-1}$, yielded slightly higher sensitivity, with the emitter exhibiting an 8% decrease in sensitivity for this single resonance. When accounting for both resonant modes of the emitter (2768 cm$^{-1}$ and 1420 cm$^{-1}$), the dual-band design exhibited a 36.5% increase in sensitivity compared to the commercial filter. When the InfraTec filter

FWHM was matched to that of the emitter (29.5 cm$^{-1}$), the emitter exhibited a 92% increase in sensitivity for the 2768 cm$^{-1}$ band alone and a 184% increase for the dual-band design. These results emphasize the tradeoff between sensitivity and selectivity in traditional NDIR sensing, and demonstrate that our a-DBR-based emitter design overcomes this limitation, enabling enhanced sensitivity without compromising spectral precision.

**CO and CO$_2$ Emitters.** In the previous section, we demonstrated that targeting multiple vibrational resonances can significantly enhance sensitivity in a filterless NDIR setup. Traditional NDIR systems often face a tradeoff between sensitivity and selectivity. Our approach overcomes this limitation by designing emitters with tailored spectral responses that achieve both. In this section, we focus on enhancing selectivity by using the same gradient descent optimization to design two high-Q emitters that independently target the stretching mode of CO ($\omega_{CO}$ = 2150 cm$^{-1}$) and asymmetric stretching mode of CO$_2$ ($\omega_{CO_2}$ = 2349 cm$^{-1}$), respectively.[37] These vibrational modes are spectrally adjacent but non-overlapping, allowing high-Q emitters to selectively detect one gas without interference or false positives from the other.

The a-DBR was deposited on highly doped CdO ($N_d$ = 3.85 × 10$^{20}$ cm$^{-3}$), whose metallic behavior in the MIR, resulting from both its negative real permittivity and non-negligible imaginary contribution, provides the near-unity reflectivity required for spectrally selective thermal emission with minimal optical losses.[44,45]

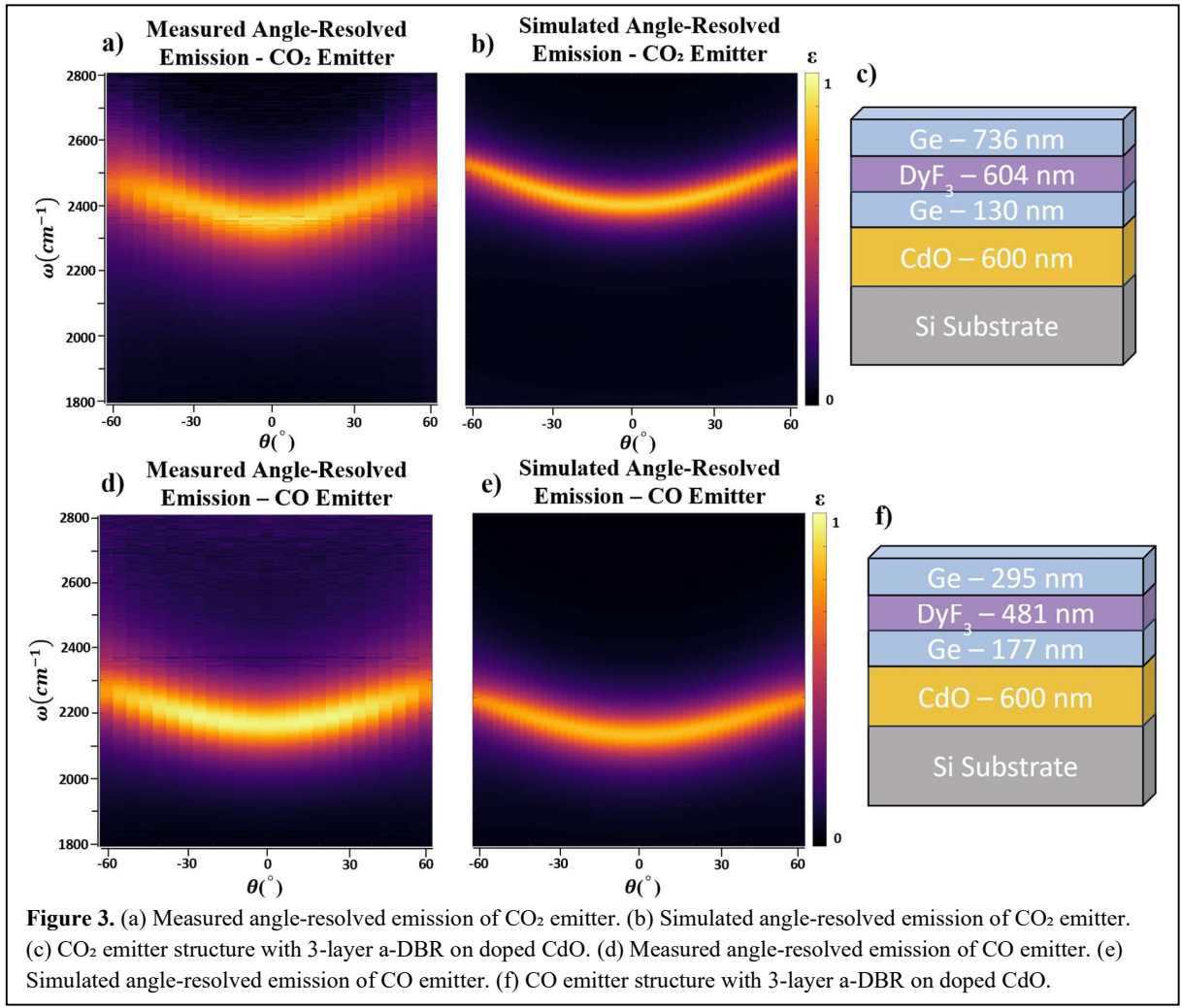

**Figure 3.** (a) Measured angle-resolved emission of CO$_2$ emitter. (b) Simulated angle-resolved emission of CO$_2$ emitter. (c) CO$_2$ emitter structure with 3-layer a-DBR on doped CdO. (d) Measured angle-resolved emission of CO emitter. (e) Simulated angle-resolved emission of CO emitter. (f) CO emitter structure with 3-layer a-DBR on doped CdO.

The high carrier density of CdO further allows tuning of its permittivity, offering additional flexibility in designing emission wavelengths.[46–51] Each CO and $CO_2$ wavelength-selective emitter consists of three alternating Ge and $DyF_3$ layers in the a-DBR **[Figs. 3(c) and 3(f)],** providing the narrow linewidths needed for selective gas detection.

Angle-resolved thermal emission measurements, performed using the previously described FTIR-based heated rotation stage, confirmed strong, narrowband emission centered at the target frequencies for both emitters. The $CO_2$-targeting emitter peaked at 2351 cm$^{-1}$ with $\varepsilon \approx 0.98$ and a Q-factor of 52.55 exhibiting high emissivity across the full $CO_2$ absorption band **[Fig. 3(a)]**. In comparison, the CO-targeting emitter peaked at 2146 cm$^{-1}$ with $\varepsilon \approx 0.98$ and a Q-factor of 76.7, similarly maintaining strong overlap with the CO absorption band **[Fig. 3(d)]**. To validate these findings, we also simulated the angular dispersion of both emitters. The $CO_2$ emitter exhibited low angular dispersion ($b_{CO_2} = 0.0806 \frac{cm^{-1}}{deg^2}$), maintaining spectral alignment with the $CO_2$ absorption band up to a collection angle of 57° **[Fig. 3(b)]**. The CO emitter showed similarly low angular dispersion ($b_{CO} = 0.0594 \frac{cm^{-1}}{deg^2}$), preserving overlap with the CO absorption band out to 60° **[Fig. 3(e)].**

Similar to the $C_3H_8$ emitter, the thermal stability of the CO and $CO_2$ emitters were also evaluated by measuring their spectral output from 200 °C to 300 °C using our FTIR system. Both emitters demonstrated minimal spectral shifting across this temperature range, with peak positions changing by only 0.23 cm$^{-1}$/°C for CO and 0.25 cm$^{-1}$/°C for $CO_2$, remaining well-aligned with their respective absorption bands (Supporting Information). These results confirm that the CO and $CO_2$ emitters maintain reliable performance under varying thermal conditions, further supporting their use in practical sensing environments.

Each emitter was again incorporated into our NDIR gas sensing setup (see Supporting Information). For $CO_2$ detection, the emitter was heated to 300 °C, and concentrations from 0 to 1,000 ppm of $CO_2$ were introduced into the gas cell. As expected, the detected power dropped in accordance with Beer-Lambert behavior **[Fig. 4(a)]**. To evaluate sensor selectivity, CO gas was flowed through the cell while the $CO_2$ emitter remained active. Despite the close spectral proximity between CO and $CO_2$ absorption bands, the pyroelectric detector registered no measurable decrease in power **[Fig. 4(b)]**. This confirms that the emitter's output exhibits minimal overlap with CO absorption features, thereby avoiding false positives and demonstrating strong spectral selectivity critical for accurate gas identification. This procedure was then repeated with the CO emitter. As CO concentrations increased, the detected power decreased **[Fig. 4(c)]**, confirming selective emission and absorption. However, a slight downward trend in detected power was observed as $CO_2$ concentration increased **[Fig. 4(d)]**. This suggests minor spectral overlap between the CO emitter output and $CO_2$ absorption features. This overlap of emissivity with other gas types can be mitigated by reducing the emission linewidth through the use of lower-loss dielectric materials or by incorporating additional a-DBR layers.

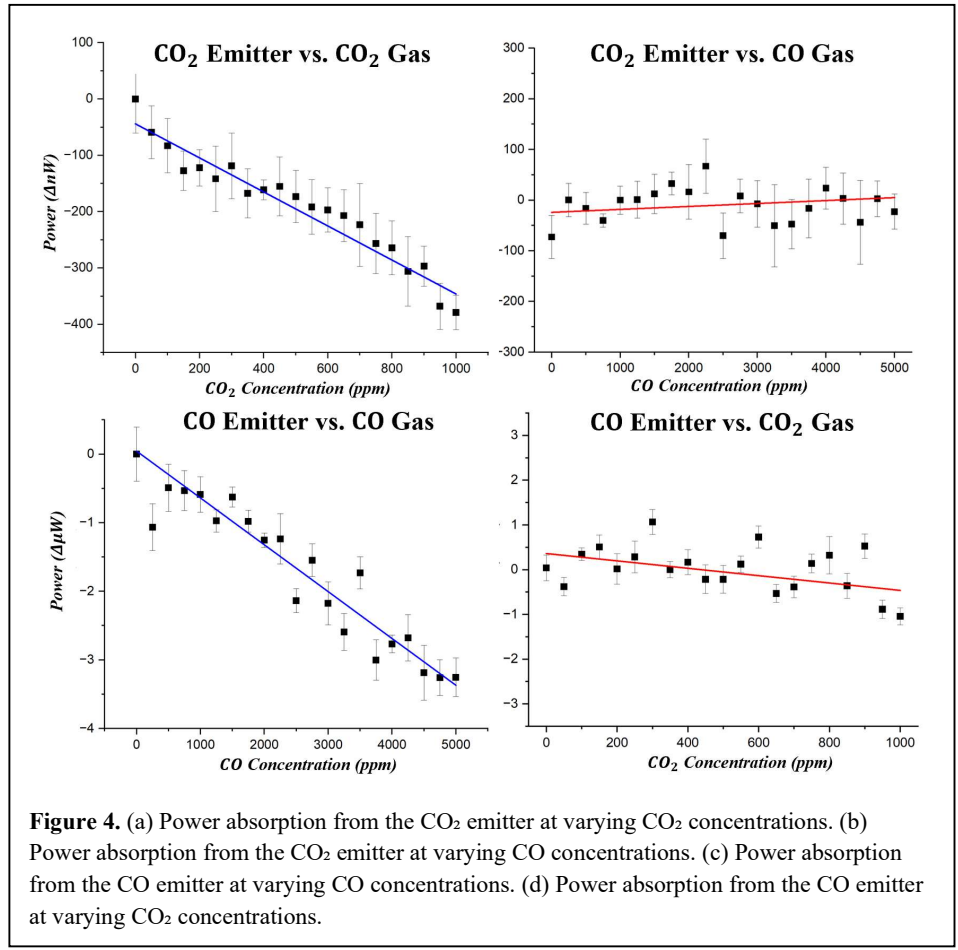

**Figure 4.** (a) Power absorption from the $CO_2$ emitter at varying $CO_2$ concentrations. (b) Power absorption from the $CO_2$ emitter at varying CO concentrations. (c) Power absorption from the CO emitter at varying CO concentrations. (d) Power absorption from the CO emitter at varying $CO_2$ concentrations.

Together, these results demonstrate that inverse-designed a-DBR emitters enable precise control over thermal emission spectra, achieving both multi-resonant and narrowband emission profiles tailored to specific molecular absorption features. The dual-band $C_3H_8$ emitter enhanced detection sensitivity by up to 184% compared to a commercial single-band NDIR filter, while the CO and $CO_2$ emitters exhibited Q-factors of 76.7 and 52.6, respectively, enabling clear spectral discrimination between closely spaced absorption features. With minimal angular dispersion, consistent performance under thermal variability, and strong selectivity in the presence of spectrally adjacent gases, the emitters demonstrate the robustness and reliability required for integration into compact, filterless NDIR gas sensors.

**Methods.**

**Emitter Fabrication.** The multi-peak thermal emitter was fabricated using radio frequency (RF) sputter deposition[52] to deposit aperiodic multilayers of Ge and $AlO_x$. Ge layers were deposited from a 2″ diameter Ge target in a pure Ar environment at a total pressure of 3 mTorr. $AlO_x$ layers were reactively sputtered from a 2″ diameter Al target in a mixed 10% $O_2$ and 90% Ar atmosphere at a total pressure of 5 mTorr. All Ge and $AlO_x$ layers were deposited sequentially at ambient temperature without breaking vacuum.

For the $C_3H_8$ emitter, the a-DBR was deposited on a low-resistivity Si substrate ($\omega_p \approx 4.73 \times 10^{12}$ rad·s$^{-1}$). For the single-peak CO and $CO_2$ emitters, an In-doped CdO layer with carrier concentration $N_d = 3.85 \times 10^{20}$ cm$^{-3}$ was first grown by reactive high-power impulse magnetron sputtering (HiPIMS)[53], following procedures described in previous reports.[35,37] The three-layer a-DBR stacks were then deposited on the CdO film at ambient temperature using electron-beam evaporation from Ge (99.999 %) and $DyF_3$ (99.9 %) sources. Layer thicknesses were monitored in real time using a quartz crystal microbalance throughout the deposition.

**Angle-Resolved Emissivity Measurements.** Angle-resolved emissivity was measured using a Bruker Vertex 70v FTIR spectrometer[54] coupled to a custom-built heated rotation stage. Spectra were collected from −60° to +60° in 5° increments with the sample temperature held between 200 and 300 °C. Angular dispersion of each emission peak was quantified by fitting the peak frequency to a second-order Taylor[55] expansion, as shown in **Eqn. 2**:

$$\omega_{r,i}(\theta) = \omega_{0,i} + \frac{b_i}{2}\theta^2, \qquad (2)$$

where $\theta$ is the emission angle, $\omega_{r,i}$ is the resonant frequency, $\omega_{0,i}$ is the emission frequency along the surface normal, and $b_i$ is the phenomenologically determined band curvature of mode $i$. The extracted $b_i$ values were used to quantify angular dispersion and evaluate spectral overlap with target absorption bands.

**Gas Sensing Measurements.** Gas sensing measurements were performed using a custom-built NDIR setup (schematic in Supporting Information). The dual-peak a-DBR emitter was mounted on a Linkam heating stage[56] and operated at 300 °C. Emission from the sample was collected using a Winston cone collimator positioned at the focal point to efficiently capture both normal and off-normal emission. The collimated beam was directed through a 10 cm gas cell (Model 162-10-CPF, Pike Technologies)[57] equipped with $CaF_2$ windows and then modulated with a mechanical chopper wheel. The modulated signal was detected with a pyroelectric detector, and the output was processed using a lock-in amplifier to monitor changes in intensity. Gas concentrations were introduced into the cell using controlled flow, and the resulting reduction in detected power was recorded and analyzed according to Beer-Lambert behavior. Additional details, including concentration calibration and signal averaging parameters are provided in the Supporting Information.

**Conclusion.** This work demonstrates a filterless NDIR sensing strategy that leverages aperiodic distributed Bragg reflectors (a-DBRs) to achieve both high sensitivity and selectivity, while minimizing fabrication complexity and cost. By applying gradient descent optimization, we designed and fabricated frequency selective thermal emitters, all without the need for lithographic patterning or additional spectral filtering components.

We first validated this approach through the development of a multi-peak emitter targeting both the C-H deformation and stretching modes of propane ($C_3H_8$). Simultaneous targeting of multiple absorption bands enabled enhanced sensitivity, particularly at low gas concentrations, and broadened the dynamic sensing range, illustrating the value of multi-resonant designs in improving detection performance. To further evaluate spectral selectivity, we developed two single-peak, high-Q emitters for carbon monoxide (CO) and carbon dioxide ($CO_2$), which possess vibrational features in close spectral proximity. Angle-resolved thermal emission measurements and gas sensing experiments confirmed strong spectral isolation and minimal crosstalk between the two targets, demonstrating the effectiveness of this approach for selective detection, even when analyte resonances are narrowly spaced.

Together, these results establish inverse-designed a-DBRs as a promising path forward for compact, low-cost NDIR gas sensors. By eliminating the need for filters, narrowband detectors, or lithographically patterned metasurfaces, this method breaks the inherent tradeoff between sensitivity and selectivity found in conventional NDIR systems. This tunable and simple design could support a wide range of emerging gas sensing applications, from environmental monitoring to medical diagnostics.

**Acknowledgements**

J.D.C. and M.H. were supported by the Office of Naval Research (ONR) under Grant No. N00014-22-1-2035, while K.D.G. was supported under ONR under Grant No. N00014-23-1-2676 and by the National Science Foundation (NSF) under Grant No. DMR-2128240. E.R.B. was supported by the Vanderbilt Innovation Catalyst Grant and E.R.B., J.R.N., C.R.G, and S.D.L. were all funded by Sensorium Technological Laboratories. J.N. gratefully acknowledges support from the Department of Defense (DoD) through the National Defense Science and Engineering Graduate (NDSEG) Fellowship Program. The authors acknowledge the use of instrumentation supported by the Vanderbilt Institute of Nanoscale Science and Engineering (VINSE) for optical and materials characterization. The authors further thank Nanotechlabs for providing the complimentary blackbody carbon nanotube sample used as a reference during FTIR thermal emission measurements.

**References**

(1) da Silveira Petruci, J. F.; da Silva Sousa, D.; Mizaikoff, B. Advanced Mid-Infrared Sensors for Molecular Analysis. *Anal. Chem.* **2025**, *97* (13), 6871–6890. https://doi.org/10.1021/acs.analchem.4c06799.
(2) Hodgkinson, J.; Tatam, R. P. Optical Gas Sensing: A Review. *Meas. Sci. Technol.* **2012**, *24* (1), 012004. https://doi.org/10.1088/0957-0233/24/1/012004.
(3) Mortazavi, S.; Makouei, S.; Abbasian, K.; Danishvar, S. Exhaled Breath Analysis (EBA): A Comprehensive Review of Non-Invasive Diagnostic Techniques for Disease Detection. *Photonics* **2025**, *12* (9), 848. https://doi.org/10.3390/photonics12090848.
(4) Flores Rangel, G.; Diaz de León Martinez, L.; Mizaikoff, B. Helicobacter Pylori Breath Test via Mid-Infrared Sensor Technology. *ACS Sens* **2025**, *10* (2), 1005–1010. https://doi.org/10.1021/acssensors.4c02785.
(5) Flores Rangel, G.; Diaz de León Martínez, L.; Walter, L. S.; Mizaikoff, B. Recent Advances and Trends in Mid-Infrared Chem/Bio Sensors. *TrAC Trends in Analytical Chemistry* **2024**, *180*, 117916. https://doi.org/10.1016/j.trac.2024.117916.
(6) Li, J.; Yu, Z.; Du, Z.; Ji, Y.; Liu, C. Standoff Chemical Detection Using Laser Absorption Spectroscopy: A Review. *Remote Sensing* **2020**, *12* (17), 2771. https://doi.org/10.3390/rs12172771.
(7) Melkonian, J.-M.; Armougom, J.; Raybaut, M.; Dherbecourt, J.-B.; Gorju, G.; Cézard, N.; Godard, A.; Pašiškevičius, V.; Coetzee, R.; Kadlčák, J. Long-Wave Infrared Multi-Wavelength Optical Source for Standoff Detection of Chemical Warfare Agents. *Appl. Opt.* **2020**, *59* (35), 11156. https://doi.org/10.1364/AO.410053.
(8) Panda, S.; Mehlawat, S.; Dhariwal, N.; Kumar, A.; Sanger, A. Comprehensive Review on Gas Sensors: Unveiling Recent Developments and Addressing Challenges. *Materials Science and Engineering: B* **2024**, *308*, 117616. https://doi.org/10.1016/j.mseb.2024.117616.
(9) Li, J.; Yu, Z.; Du, Z.; Ji, Y.; Liu, C. Standoff Chemical Detection Using Laser Absorption Spectroscopy: A Review. *Remote Sensing* **2020**, *12* (17), 2771. https://doi.org/10.3390/rs12172771.
(10) Todd, L. Measuring Chemical Emissions Using Open-Path Fourier Transform Infrared (OP-FTIR) Spectroscopy and Computer-Assisted Tomography. *Atmospheric Environment* **2001**, *35* (11), 1937–1947. https://doi.org/10.1016/S1352-2310(00)00546-X.
(11) Method TO-16 - Long-Path Open-Path Fourier Transform Infrared Monitoring Of Atmospheric Gases.
(12) Schill, S.; McEwan, R. S.; Moffet, R.; Marrero, J.; MacDonald, C.; Winegar, E. Real-World Application of Open-Path UV-DOAS, TDL, and FT-IR Spectroscopy for Air Quality Monitoring at Industrial Facilities. **2022**, *37*, 18–22.
(13) Saggin, B.; Comolli, L.; Formisano, V. Mechanical Disturbances in Fourier Spectrometers. *Appl. Opt., AO* **2007**, *46* (22), 5248–5256. https://doi.org/10.1364/AO.46.005248.
(14) Comolli, L.; Saggin, B. Evaluation of the Sensitivity to Mechanical Vibrations of an IR Fourier Spectrometer. *Rev. Sci. Instrum.* **2005**, *76* (12), 123112. https://doi.org/10.1063/1.2149009.

(15) Casasanta, G.; Garra, R. Towards a Generalized Beer-Lambert Law. *Fractal and Fractional* **2018**, *2* (1), 8. https://doi.org/10.3390/fractalfract2010008.

(16) Gandhi, K.; Sharma, N.; Gautam, P. B.; Sharma, R.; Mann, B.; Pandey, V. Spectroscopy. In *Advanced Analytical Techniques in Dairy Chemistry*; Springer, New York, NY, 2022; pp 161–176. https://doi.org/10.1007/978-1-0716-1940-7_8.

(17) Koppens, F. H. L.; Mueller, T.; Avouris, P.; Ferrari, A. C.; Vitiello, M. S.; Polini, M. Photodetectors Based on Graphene, Other Two-Dimensional Materials and Hybrid Systems. *Nature Nanotech* **2014**, *9* (10), 780–793. https://doi.org/10.1038/nnano.2014.215.

(18) Meyer, J. R.; Vurgaftman, I.; Canedy, C. L.; Bewley, W. W.; Kim, C. S.; Merritt, C. D.; Warren, M. V.; Kim, M. In-Plane Resonant-Cavity Infrared Photodetectors with Fully-Depleted Absorbers. US10559704B2, February 11, 2020. https://patents.google.com/patent/US10559704B2/en (accessed 2025-05-22).

(19) Folland, T. G.; Nordin, L.; Wasserman, D.; Caldwell, J. D. Probing Polaritons in the Mid- to Far-Infrared. *Journal of Applied Physics* **2019**, *125* (19), 191102. https://doi.org/10.1063/1.5090777.

(20) Krier, A.; Repiso, E.; Al-Saymari, F.; Carrington, P. J.; Marshall, A. R. J.; Qi, L.; Krier, S. E.; Lulla, K. J.; Steer, M.; MacGregor, C.; Broderick, C. A.; Arkani, R.; O'Reilly, E.; Sorel, M.; Molina, S. I.; De La Mata, M. 2 - Mid-Infrared Light-Emitting Diodes. In *Mid-infrared Optoelectronics*; Tournié, E., Cerutti, L., Eds.; Woodhead Publishing Series in Electronic and Optical Materials; Woodhead Publishing, 2020; pp 59–90. https://doi.org/10.1016/B978-0-08-102709-7.00002-4.

(21) Al-Saymari, F. A.; Craig, A. P.; Noori, Y. J.; Lu, Q.; Marshall, A. R. J.; Krier, A. Electroluminescence Enhancement in Mid-Infrared InAsSb Resonant Cavity Light Emitting Diodes for CO2 Detection. *Applied Physics Letters* **2019**, *114* (17), 171103. https://doi.org/10.1063/1.5090840.

(22) De Zoysa, M.; Asano, T.; Mochizuki, K.; Oskooi, A.; Inoue, T.; Noda, S. Conversion of Broadband to Narrowband Thermal Emission through Energy Recycling. *Nature Photon* **2012**, *6* (8), 535–539. https://doi.org/10.1038/nphoton.2012.146.

(23) Laroche, M.; Carminati, R.; Greffet, J.-J. Coherent Thermal Antenna Using a Photonic Crystal Slab. *Phys. Rev. Lett.* **2006**, *96* (12), 123903. https://doi.org/10.1103/PhysRevLett.96.123903.

(24) Inoue, T.; De Zoysa, M.; Asano, T.; Noda, S. High-Q Mid-Infrared Thermal Emitters Operating with High Power-Utilization Efficiency. *Opt. Express* **2016**, *24* (13), 15101. https://doi.org/10.1364/oe.24.015101.

(25) Lu, G.; Nolen, J. R.; Folland, T. G.; Tadjer, M. J.; Walker, D. G.; Caldwell, J. D. Narrowband Polaritonic Thermal Emitters Driven by Waste Heat. *ACS Omega* **2020**, *5* (19), 10900–10908. https://doi.org/10.1021/acsomega.0c00600.

(26) Greffet, J.-J.; Carminati, R.; Joulain, K.; Mulet, J.-P.; Mainguy, S.; Chen, Y. Coherent Emission of Light by Thermal Sources. *Nature* **2002**, *416* (6876), 61–64. https://doi.org/10.1038/416061a.

(27) Wang, T.; Li, P.; Chigrin, D. N.; Giles, A. J.; Bezares, F. J.; Glembocki, O. J.; Caldwell, J. D.; Taubner, T. Phonon-Polaritonic Bowtie Nanoantennas: Controlling Infrared Thermal Radiation at the Nanoscale. *ACS Photonics* **2017**, *4* (7), 1753–1760. https://doi.org/10.1021/acsphotonics.7b00321.

(28) Schuller, J. A.; Taubner, T.; Brongersma, M. L. Optical Antenna Thermal Emitters. *Nature Photon* **2009**, *3* (11), 658–661. https://doi.org/10.1038/nphoton.2009.188.

(29) Nolen, J. R.; Overvig, A. C.; Cotrufo, M.; Alù, A. Local Control of Polarization and Geometric Phase in Thermal Metasurfaces. *Nat. Nanotechnol.* **2024**, *19* (11), 1627–1634. https://doi.org/10.1038/s41565-024-01763-6.

(30) Livingood, A.; Nolen, J. R.; Folland, T. G.; Potechin, L.; Lu, G.; Criswell, S.; Maria, J.-P.; Shelton, C. T.; Sachet, E.; Caldwell, J. D. Filterless Nondispersive Infrared Sensing Using Narrowband Infrared Emitting Metamaterials. *ACS Photonics* **2021**, *8* (2), 472–480. https://doi.org/10.1021/acsphotonics.0c01432.

(31) Howes, A.; Nolen, J. R.; Caldwell, J. D.; Valentine, J. Near-Unity and Narrowband Thermal Emissivity in Balanced Dielectric Metasurfaces. *Advanced Optical Materials* **2020**, *8* (4), 1901470. https://doi.org/10.1002/adom.201901470.

(32) Mason, J. A.; Smith, S.; Wasserman, D. Strong Absorption and Selective Thermal Emission from a Midinfrared Metamaterial. *Applied Physics Letters* **2011**, *98* (24), 241105. https://doi.org/10.1063/1.3600779.

(33) Overvig, A. C.; Mann, S. A.; Alù, A. Thermal Metasurfaces: Complete Emission Control by Combining Local and Nonlocal Light-Matter Interactions. *Phys. Rev. X* **2021**, *11* (2), 021050. https://doi.org/10.1103/PhysRevX.11.021050.

(34) Caldwell, J. D.; Lindsay, L.; Giannini, V.; Vurgaftman, I.; Reinecke, T. L.; Maier, S. A.; Glembocki, O. J. Low-Loss, Infrared and Terahertz Nanophotonics Using Surface Phonon Polaritons. *Nanophotonics* **2015**, *4* (1), 44–68. https://doi.org/10.1515/nanoph-2014-0003.

(35) He, M.; Nolen, J. R.; Nordlander, J.; Cleri, A.; Lu, G.; Arnaud, T.; McIlwaine, N. S.; Diaz-Granados, K.; Janzen, E.; Folland, T. G.; Edgar, J. H.; Maria, J.-P.; Caldwell, J. D. Coupled Tamm Phonon and Plasmon Polaritons for Designer Planar Multiresonance Absorbers. *Advanced Materials* **2023**, *35* (20), 2209909. https://doi.org/10.1002/adma.202209909.

(36) Yang, Z.-Y.; Ishii, S.; Yokoyama, T.; Dao, T. D.; Sun, M.-G.; Pankin, P. S.; Timofeev, I. V.; Nagao, T.; Chen, K.-P. Narrowband Wavelength Selective Thermal Emitters by Confined Tamm Plasmon Polaritons. *ACS Photonics* **2017**, *4* (9), 2212–2219. https://doi.org/10.1021/acsphotonics.7b00408.

(37) He, M.; Nolen, J. R.; Nordlander, J.; Cleri, A.; McIlwaine, N. S.; Tang, Y.; Lu, G.; Folland, T. G.; Landman, B. A.; Maria, J.-P.; Caldwell, J. D. Deterministic Inverse Design of Tamm Plasmon Thermal Emitters with Multi-Resonant Control. *Nat. Mater.* **2021**, *20* (12), 1663–1669. https://doi.org/10.1038/s41563-021-01094-0.

(38) Wang, Z.; Clark, J. K.; Ho, Y.-L.; Vilquin, B.; Daiguji, H.; Delaunay, J.-J. Narrowband Thermal Emission from Tamm Plasmons of a Modified Distributed Bragg Reflector. *Applied Physics Letters* **2018**, *113* (16), 161104. https://doi.org/10.1063/1.5048950.

(39) Informatics, N. O. of D. and. *Propane*. https://webbook.nist.gov/cgi/cbook.cgi?ID=C74986&Type=IR-SPEC&Index=1 (accessed 2025-06-02).

(40) Informatics, N. O. of D. and. *Carbon dioxide*. https://webbook.nist.gov/cgi/cbook.cgi?ID=C124389&Type=IR-SPEC&Index=1 (accessed 2024-10-30).

(41) Informatics, N. O. of D. and. *Carbon monoxide*. https://webbook.nist.gov/cgi/cbook.cgi?ID=C630080&Type=IR-SPEC&Index=1 (accessed 2024-10-30).

(42) Passler, N. C.; Paarmann, A. Generalized 4 × 4 Matrix Formalism for Light Propagation in Anisotropic Stratified Media: Study of Surface Phonon Polaritons in Polar Dielectric Heterostructures. *J. Opt. Soc. Am. B, JOSAB* **2017**, *34* (10), 2128–2139. https://doi.org/10.1364/JOSAB.34.002128.

(43) *Get IR filters and IR windows from InfraTec*. https://www.infratec-infrared.com/sensor-division/ir-filters/ (accessed 2025-10-06).

(44) Nolen, J. R.; Runnerstrom, E. L.; Kelley, K. P.; Luk, T. S.; Folland, T. G.; Cleri, A.; Maria, J.-P.; Caldwell, J. D. Ultraviolet to Far-Infrared Dielectric Function of $n$-Doped Cadmium Oxide Thin Films. *Phys. Rev. Mater.* **2020**, *4* (2), 025202. https://doi.org/10.1103/PhysRevMaterials.4.025202.

(45) Sachet, E.; Shelton, C. T.; Harris, J. S.; Gaddy, B. E.; Irving, D. L.; Curtarolo, S.; Donovan, B. F.; Hopkins, P. E.; Sharma, P. A.; Sharma, A. L.; Ihlefeld, J.; Franzen, S.; Maria, J.-P. Dysprosium-Doped Cadmium Oxide as a Gateway Material for Mid-Infrared Plasmonics. *Nature Mater* **2015**, *14* (4), 414–420. https://doi.org/10.1038/nmat4203.

(46) Nolen, J. R.; Runnerstrom, E. L.; Kelley, K. P.; Luk, T. S.; Folland, T. G.; Cleri, A.; Maria, J.-P.; Caldwell, J. D. Ultraviolet to Far-Infrared Dielectric Function of $n$-Doped Cadmium Oxide Thin Films. *Phys. Rev. Mater.* **2020**, *4* (2), 025202. https://doi.org/10.1103/PhysRevMaterials.4.025202.

(47) Kelley, K. P.; Sachet, E.; Shelton, C. T.; Maria, J.-P. High Mobility Yttrium Doped Cadmium Oxide Thin Films. *APL Materials* **2017**, *5* (7), 076105. https://doi.org/10.1063/1.4993799.

(48) Runnerstrom, E. L.; Kelley, K. P.; Sachet, E.; Shelton, C. T.; Maria, J.-P. Epsilon-near-Zero Modes and Surface Plasmon Resonance in Fluorine-Doped Cadmium Oxide Thin Films. *ACS Photonics* **2017**, *4* (8), 1885–1892. https://doi.org/10.1021/acsphotonics.7b00429.

(49) Runnerstrom, E. L.; Kelley, K. P.; Folland, T. G.; Nolen, J. R.; Engheta, N.; Caldwell, J. D.; Maria, J.-P. Polaritonic Hybrid-Epsilon-near-Zero Modes: Beating the Plasmonic Confinement vs Propagation-Length Trade-Off with Doped Cadmium Oxide Bilayers. *Nano Lett.* **2019**, *19* (2), 948–957. https://doi.org/10.1021/acs.nanolett.8b04182.

(50) Cleri, A.; Tomko, J.; Quiambao-Tomko, K.; Imperatore, M. V.; Zhu, Y.; Nolen, J. R.; Nordlander, J.; Caldwell, J. D.; Mao, Z.; Giebink, N. C.; Kelley, K. P.; Runnerstrom, E. L.; Hopkins, P. E.; Maria, J.-P. Mid-Wave to near-IR Optoelectronic Properties and Epsilon-near-Zero Behavior in Indium-Doped Cadmium Oxide. *Phys. Rev. Mater.* **2021**, *5* (3), 035202. https://doi.org/10.1103/PhysRevMaterials.5.035202.

(51) Runnerstrom, E. L.; Kelley, K. P.; Folland, T. G.; Engheta, N.; Caldwell, J. D.; Maria, J.-P. Polaritonic Hybrid-Epsilon-near-Zero Modes: Engineering Strong Optoelectronic Coupling and Dispersion in Doped Cadmium Oxide Bilayers. arXiv August 11, 2018. https://doi.org/10.48550/arXiv.1808.03847.

(52) Raven, M. S. Radio Frequency Sputtering and the Deposition of High-Temperature Superconductors. *J Mater Sci: Mater Electron* **1994**, *5* (3), 129–146. https://doi.org/10.1007/BF01198944.

(53) Borowski, P.; Myśliwiec, J. Recent Advances in Magnetron Sputtering: From Fundamentals to Industrial Applications. *Coatings* **2025**, *15* (8), 922. https://doi.org/10.3390/coatings15080922.

(54) VERTEX_70-HYPERION_2000_22.Pdf. https://www.depts.ttu.edu/coe/research/mcc/documents/VERTEX_70-HYPERION_2000_22.pdf (accessed 2025-06-02).

(55) Duan, W.; Chen, J.; Zhao, B. Second-Order Taylor Expansion Boundary Element Method for the Second-Order Wave Diffraction Problem. *Engineering Analysis with Boundary Elements* **2015**, *58*, 140–150. https://doi.org/10.1016/j.enganabound.2015.04.008.

(56) *Linkam THMS600 - Temperature Control Stage for Microscopy and Spectroscopy*. Linkam Scientific. https://www.linkam.co.uk/thms600 (accessed 2025-06-02).

(57) PIKE-Technologies_Stainless-Steel-Short-Path-Gas-Cells.Pdf. https://www.piketech.com/wp-content/uploads/PDS/transmission/PIKE-Technologies_Stainless-Steel-Short-Path-Gas-Cells.pdf (accessed 2025-09-08).

# Supplementary Information

# Multi-resonant non-dispersive infrared gas sensing: breaking the selectivity and sensitivity tradeoff

Emma R. Bartelsen[1,2], J. Ryan Nolen[3], Christopher R. Gubbin[3], Mingze He[2,4], Ryan W. Spangler[5], Joshua Nordlander[5], Katja Diaz-Granados[1], Simone De Liberato[3,6,7], Jon-Paul Maria[5], James R. McBride[8], Joshua D. Caldwell[1,2,3]

[1] Interdisciplinary Materials Science Program, Vanderbilt University, Nashville, Tennessee 37240, USA
[2] Department of Mechanical Engineering, Vanderbilt University, Nashville, Tennessee 37235, USA
[3] Sensorium Technological Labs, 6714 Duquaine Ct, Nashville, Tennessee 37205, USA
[4] Photonics Initiative, Advanced Science Research Center, City University of New York, New York, NY 10031, USA
[5] Department of Materials Science and Engineering, The Pennsylvania State University, University Park, Pennsylvania 16802, USA
[6] Istituto di Fotonica e Nanotecnologie, Consiglio Nazionale delle Ricerche (CNR), Piazza Leonardo da Vinci 32, Milano, 20133, Italy
[7] School of Physics and Astronomy, University of Southampton, University Road, Southampton, SO17 1BJ, United Kingdom
[8] Department of Chemistry, The Vanderbilt Institute of Nanoscale Science and Engineering, Vanderbilt University, Nashville, TN 37235, USA

## S.1 Temporal Coupled Mode Theory (TCMT)

Temporal coupled mode theory (TCMT) is a compact analytical framework used to describe how electromagnetic waves couple into and out of resonant modes in photonic structures.[1] In this approach, the resonant response is modeled by Lorentzian functions defined by the central frequency, linewidth, and the coupling strength to external channels. TCMT links the temporal dynamics of resonant modes with their absorptivity and emissivity spectra, making it well-suited for describing systems such as distributed Bragg reflector cavities, plasmonic resonances, and Tamm plasmon modes.

Here, TCMT was used to model the thermal emission spectra of the designed Tamm plasmon emitters and to estimate the power collected by the parabolic mirror in the experimental setup. The emission profiles were represented as Lorentzian resonances centered at frequencies extracted from experimental measurements, with linewidth and curvature parameters determined by fitting. For the CO and $CO_2$ emitters, a single Lorentzian resonance was sufficient, while the $C_3H_8$ emitter required two resonances to capture its multiple vibrational modes. This framework enabled calculation of angle- and frequency-dependent emissivity, incorporation of Planck's blackbody distribution at the emitter temperature, and angular integration to estimate the total collected power. These results provided a physically meaningful way to connect experimental measurements with modeled emission spectra.

## S.2 Stochastic gradient descent implementation for Tamm plasmon design

A stochastic gradient descent (SGD) inverse design framework, previously developed and published by our group[2], was used to optimize the distributed Bragg reflector (DBR) stacks supporting the Tamm plasmon resonances in this work. The framework integrates a differentiable transfer matrix method with the Adam optimizer (Adaptive Moment Estimation) to adjust layer thicknesses until the calculated resonance matches a user-defined spectral target.

The framework was applied to design multilayer Ge/DyF$_3$ stacks on doped CdO substrates for CO and CO$_2$ detection, and AlO$_x$/Ge stacks on Si substrates for C$_3$H$_8$ emission. Each design targeted narrowband transmission features centered at the vibrational frequencies of the corresponding gas. The algorithm outputs the optimized thickness profile, designed spectrum, and convergence history, which were subsequently compared with experimental measurements. The full implementation and example training scripts are available in Ref. 2[3], and a detailed description of the methodology is provided in Ref. 1[2].

## S.3 Variable angular dispersion modeling via Taylor expansion

The angular dispersion of the Tamm plasmon resonances was modeled using a second-order Taylor expansion around normal incidence ($\theta = 0$):

$$\omega_{r,i}(\theta) \approx \omega_{0,i} + \frac{1}{2} b_i \theta^2$$

Where $\omega_{r,i}$ is the resonance frequency at emission angle $\theta$, $\omega_{0,i}$ is the resonance frequency at normal incidence, and $b_i$ is the band curvature of mode $i$. This expansion provides an approximation of the resonance dispersion, reducing the angular dependence to a single curvature parameter. This form is well-suited to our emitters, as the TP modes exhibit smooth, symmetric dispersion about the surface normal. The fitted curvature values reported in the main text thus directly quantify the robustness of spectral overlap with C$_3$H$_8$, CO$_2$, and CO absorption across a broad angular range.

## S.4 Dielectric functions of Ge, AlO$_x$, and DyF$_3$ at 300 °C

The dielectric functions of Ge, AlO$_x$, and DyF$_3$ at 300 °C were used as input parameters for transfer matrix simulations of the a-DBR stacks. Experimental ellipsometry provided the tabulated datasets, giving the real ($\varepsilon'$) and imaginary ($\varepsilon''$) parts of the permittivity as a function of wavenumber. From these values, the complex refractive index $\tilde{n} = n + ik = \sqrt{\varepsilon' + i\varepsilon''}$ was calculated and incorporated into the optimization and design framework.

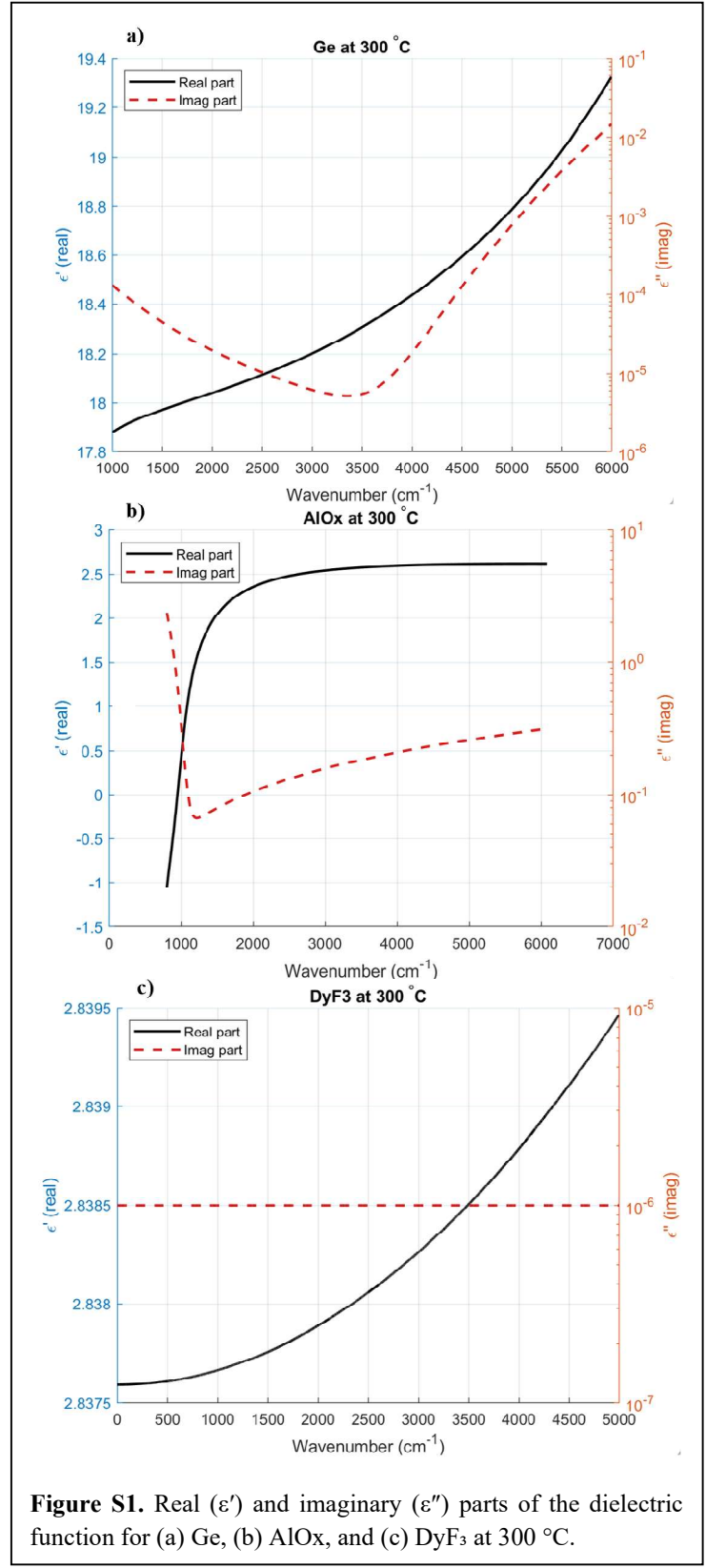

**Figure S1.** Real ($\varepsilon'$) and imaginary ($\varepsilon''$) parts of the dielectric function for (a) Ge, (b) AlOx, and (c) DyF$_3$ at 300 °C.

For Ge, the dielectric response across the mid-infrared is weakly dispersive, with $\varepsilon' \approx 18 - 19$ and $\varepsilon'' \lesssim 10^{-2}$. This corresponds to a high refractive index (n ≈ 4.2) with negligible absorption (k ≈ 0), consistent with Ge being a low-loss, high-index material in this spectral window. In the CO and $CO_2$ operation windows, from 2100 to 2400 cm$^{-1}$, $DyF_3$ exhibits $\varepsilon' \approx 2.838$ with $\varepsilon'' \approx 0$, which results in n ≈ 1.68 and k ≈ 0. In the propane window near 2700 to 3000 cm$^{-1}$, $AlO_x$ exhibits $\varepsilon'$ approximately 2.3 to 2.5 with small $\varepsilon''$, which corresponds to n ≈ 1.5 to 1.6 and k ≈ 0. $DyF_3$ and $AlO_x$ therefore serve as the low-index counterpart to Ge in their respective bands, with index contrasts relative to Ge of Δn ≈ 2.5 for $DyF_3$ and Δn ≈ 2.6 to 2.7 for $AlO_x$, which is sufficient to open wide photonic bandgaps and to support narrowband Tamm plasmon modes.

The selection between $DyF_3$ and $AlO_x$ reflects a balance between optical transparency and fabrication practicality. $DyF_3$ is more ideal optically because it is extremely transparent in the mid-infrared, but precise control of layer thickness is more difficult to achieve during sputtering. The CO and CO$_2$ emitters consisted of four layers arranged as CdO, Ge, DyF$_3$ and Ge. Using DyF$_3$ in for these a-DBR stacks was practical because only a single DyF$_3$ layer was required. The dual emitter for propane required an eight-layer stack. In this case, using multiple layers of DyF$_3$ would have introduced greater cumulative error, since achieving nanometer-level thickness control with DyF$_3$ is more difficult than with AlO$_x$. To minimize the accumulation of deviations across the stack, AlO$_x$ was selected as the low-index counterpart, prioritizing deposition reliability even though its losses are slightly higher.

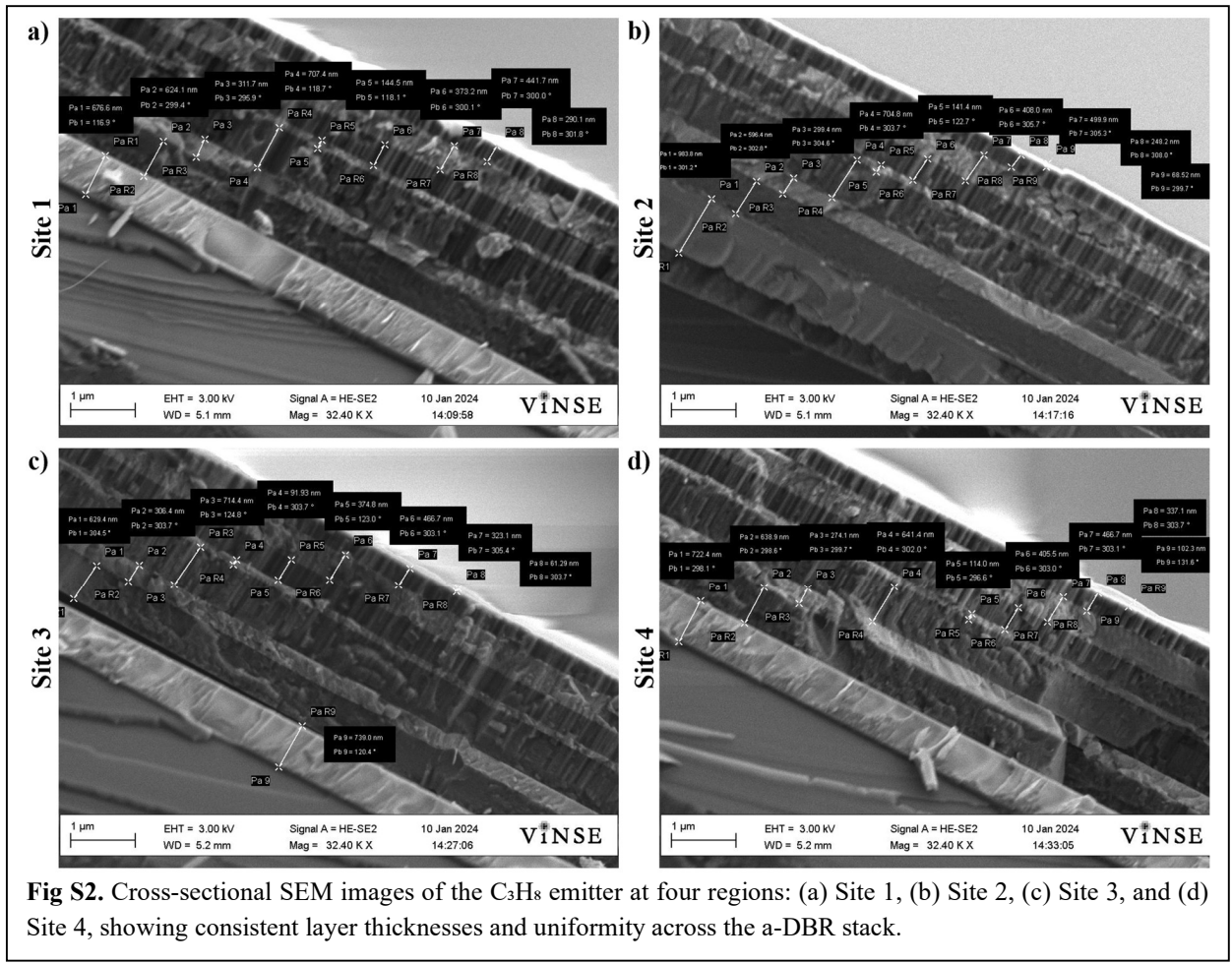

**Fig S2.** Cross-sectional SEM images of the $C_3H_8$ emitter at four regions: (a) Site 1, (b) Site 2, (c) Site 3, and (d) Site 4, showing consistent layer thicknesses and uniformity across the a-DBR stack.

**S.5 Target and as-grown layer thicknesses of a-DBR stacks**

Accurate control of layer thickness is critical for achieving the designed spectral response of each stack, as small deviations can shift resonance positions or broaden the emission linewidths. To assess the fidelity and uniformity of the deposition process, cross-sectional SEM measurements were taken at multiple points across each structure. Values from four regions (**Figs. S2 and S4**) were measured and averaged to account for spatial variations across the wafer. From these measurements, the mean, median, standard deviation, and percent error were calculated relative to the target design (**Tables S2 and S3**), providing a quantitative evaluation of growth precision. Overall, the layer stacks closely match their target thicknesses, confirming uniform deposition, consistent material growth rates, and reproducible fabrication across samples.

**S.5.1 C$_3$H$_8$ a-DBR**

The C$_3$H$_8$ emitter consisted of eight alternating Ge and AlO$_x$ layers deposited on Si. Cross-sectional SEM images from four regions of the wafer (**Figs. S2a – S2d**) were used to measure individual layer thicknesses. The expected and measured values are summarized in **Table S5**, with statistical results shown in **Fig. S3**.

**Table S1.** Expected and measured layer thicknesses for the C$_3$H$_8$ emitter obtained from SEM cross-sections at four regions (Site 1 – 4). Reported values include the mean, median, standard deviation (SD), absolute error, and percent error relative to the target design.

| C$_3$H$_8$ a-DBR Layer Thicknesses (nm) | | | | | | | | | | |
|---|---|---|---|---|---|---|---|---|---|---|
| *Layer #* | *Expected* | *Site 1* | *Site 2* | *Site 3* | *Site 4* | *Mean* | *Median* | *SD* | *Absolute Error (nm)* | *Percent Error (%)* |
| 1 | 698 | 676.6 | 983.8 | 739.0 | 722.4 | 780.5 | 730.7 | 138.1 | 82.45 | 11.81 |
| 2 | 612 | 624.1 | 596.4 | 629.4 | 638.9 | 622.2 | 626.8 | 18.26 | 10.20 | 1.667 |
| 3 | 303 | 311.7 | 299.4 | 306.4 | 274.1 | 297.9 | 302.9 | 16.65 | 5.100 | -1.683 |
| 4 | 670 | 707.4 | 704.8 | 714.4 | 641.4 | 692.0 | 706.1 | 33.98 | 22.00 | 3.284 |
| 5 | 108 | 144.5 | 141.4 | 91.93 | 114.0 | 123.0 | 127.7 | 24.81 | 14.96 | 13.85 |
| 6 | 383 | 373.2 | 408.0 | 374.8 | 405.5 | 390.4 | 390.2 | 18.95 | 7.375 | 1.926 |
| 7 | 453 | 441.7 | 499.9 | 466.7 | 466.7 | 468.8 | 466.7 | 23.88 | 15.75 | 3.477 |
| 8 | 297 | 290.1 | 248.2 | 323.1 | 337.1 | 299.6 | 306.6 | 39.54 | 2.625 | 0.884 |
| Average | | | | | | | | 39.27 | 20.06 | 4.402 |

Across all layers, the mean deviation from the design was 4.4%, with the largest variation (13.9%) occurring in the thinnest Ge layer (≈ 100 nm), which is more sensitive to rate fluctuations during deposition. The average absolute error was 39.27 nm, demonstrating close agreement between designed and fabricated values. These results confirm that the layer thickness precision achieved is sufficient to maintain the intended resonance behavior of the structure.

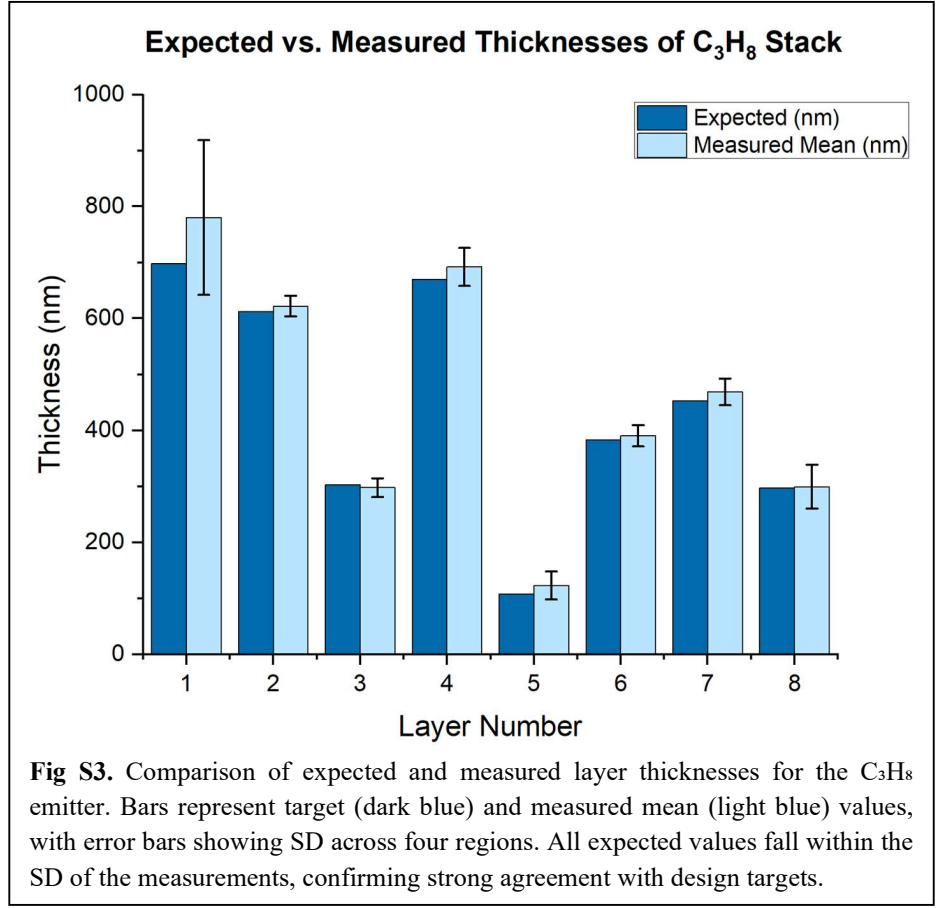

**Fig S3.** Comparison of expected and measured layer thicknesses for the $C_3H_8$ emitter. Bars represent target (dark blue) and measured mean (light blue) values, with error bars showing SD across four regions. All expected values fall within the SD of the measurements, confirming strong agreement with design targets.

### S.5.2 CO a-DBR

The CO and $CO_2$ emitters were grown on a doped CdO substrate, creating challenges in SEM imaging due to surface charging under the electron beam. To mitigate this, a 100 nm Au overlayer was deposited by resistive thermal evaporation to provide a conductive path and reduce charging. Prior to cross-sectioning, each region of interest was capped with a Pt layer deposited in situ using electron-beam-assisted deposition to protect the surface during focused ion beam (FIB) milling. The cross-sections were then prepared using a $Ga^+$ ion beam to expose the layer structure for high-resolution SEM imaging (**Fig. S4**).

SEM cross-sections from four regions (Figs. S4a–S4d) were analyzed to extract as-grown layer thicknesses. The results are summarized in **Table S6** and plotted in **Fig. S5**. The CO emitter exhibited an average percent error of 11.5%, with the smallest deviation (1.6%) in the ≈ 480 nm $AlO_x$ layer and the largest (24.4%) in the ≈ 600 nm Ge layer. The average absolute error was ≈ 50 nm. Despite these deviations, the relative thickness ratios remained within the tolerances required to maintain the intended optical interference and spectral response. The $CO_2$ emitter was processed and analyzed using the same methodology, and the results reflected similar trends to those observed for the CO emitter.

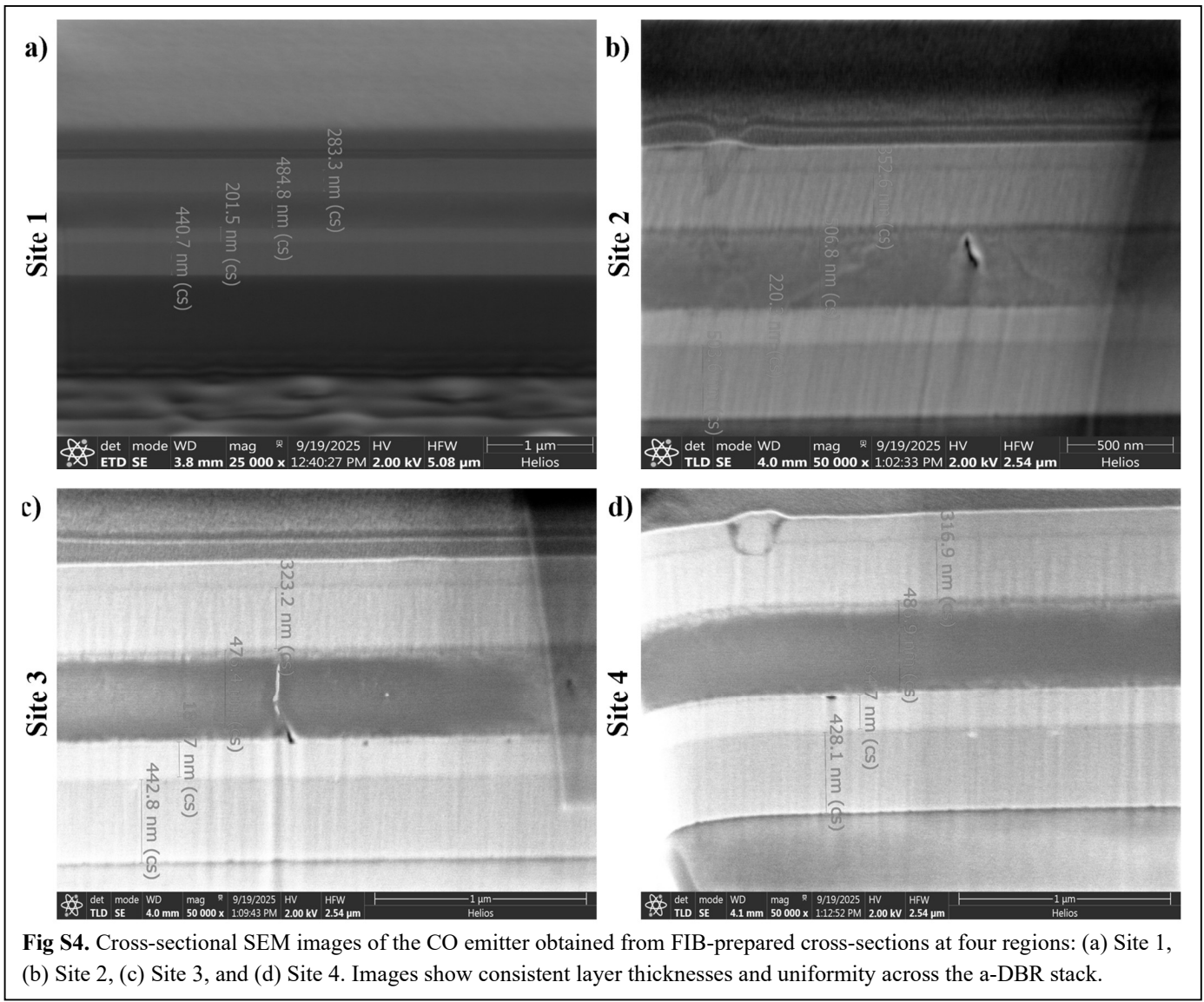

**Fig S4.** Cross-sectional SEM images of the CO emitter obtained from FIB-prepared cross-sections at four regions: (a) Site 1, (b) Site 2, (c) Site 3, and (d) Site 4. Images show consistent layer thicknesses and uniformity across the a-DBR stack.

Although the CdO layer deviated from its target thickness, this variation does not significantly affect device performance. Once the CdO film exceeds approximately 300 nm, it behaves as an optically thick, highly reflective substrate in the mid-infrared. At this thickness, the electromagnetic field penetration depth is much smaller than the film thickness, and the substrate effectively acts as a semi-infinite reflector. Thus, moderate deviations in CdO thickness do not alter the resonance condition or overall emission characteristics.

**Table S2.** Expected and measured layer thicknesses for the CO emitter obtained from FIB-SEM cross-sections at four regions (Site 1–4). Reported values include the mean, median, standard deviation (SD), absolute error, and percent error relative to the target design.

**CO a-DBR Layer Thicknesses (nm)**

| Layer # | Expected | Site 1 | Site 2 | Site 3 | Site 4 | Mean | Median | SD | Absolute Error (nm) | Percent Error (%) |
|---|---|---|---|---|---|---|---|---|---|---|
| 1 | 295 | 283.3 | 352.6 | 323.2 | 316.9 | 319.0 | 320.1 | 28.43 | 24.00 | 8.14 |
| 2 | 481 | 484.8 | 506.8 | 476.4 | 486.9 | 488.7 | 485.9 | 18.26 | 7.73 | 1.61 |
| 3 | 177 | 201.5 | 220.3 | 184.7 | 184.7 | 197.8 | 193.1 | 16.65 | 20.80 | 11.75 |
| 4 | 600 | 440.7 | 503.6 | 442.8 | 428.1 | 453.8 | 441.8 | 33.98 | 146.20 | 24.37 |
| Average | | | | | | | | 23.03 | 49.68 | 11.46 |

### S.6 Band-Dependent Contributions to Propane Detection

Simulated transmission spectra were generated to evaluate how the two emission bands of the dual-band $C_3H_8$ emitter contribute to propane sensing across varying gas concentrations. **Figure S8** presents a false-color map of the simulated spectral response as a function of wavenumber and propane concentration, where yellow indicates stronger absorption and blue corresponds to minimal absorption or higher transmission.

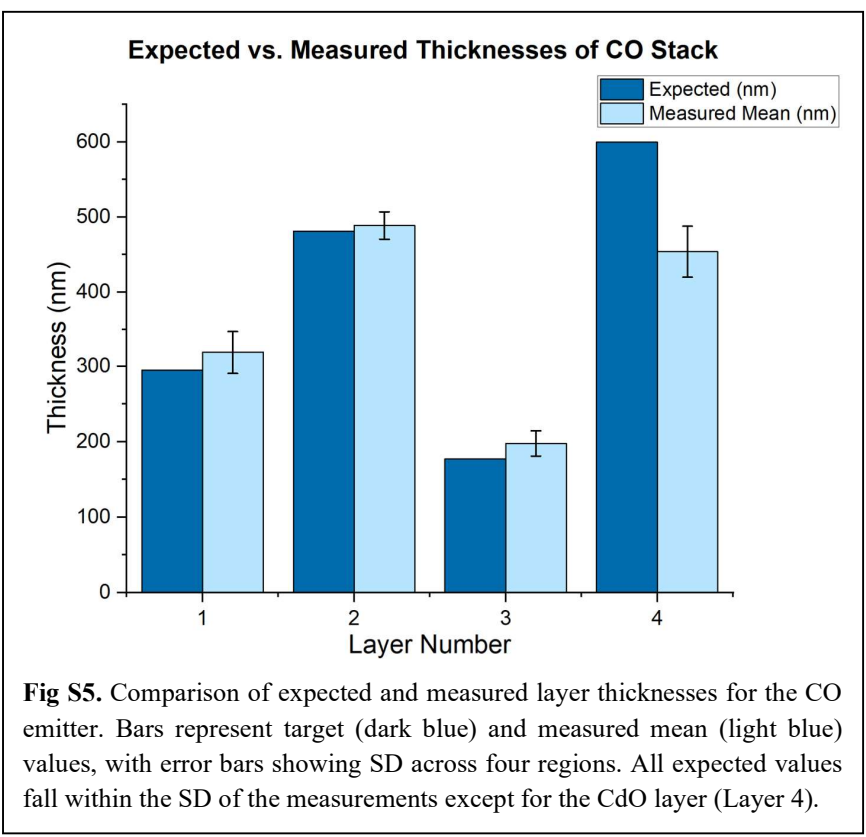

**Fig S5.** Comparison of expected and measured layer thicknesses for the CO emitter. Bars represent target (dark blue) and measured mean (light blue) values, with error bars showing SD across four regions. All expected values fall within the SD of the measurements except for the CdO layer (Layer 4).

Two primary absorption features are observed near 2768 cm$^{-1}$ and 1420 cm$^{-1}$, corresponding to the C–H stretching and C–H bending vibrational modes of propane, respectively. As the propane concentration increases from 0% to 100%, both bands exhibit progressively stronger absorption, consistent with increased interaction between the emitted light and the target gas. However, each resonance displays a distinct sensitivity profile with concentration.

At low propane concentrations (below approximately 30%), the upper band at 2768 cm$^{-1}$ dominates the response due to its higher absorption coefficient, producing a rapid increase in absorption intensity. As concentration increases beyond this point, this band begins to saturate, where additional propane produces minimal further absorption. The lower band at 1420 cm$^{-1}$, however, continues to vary with concentration, maintaining a measurable response even after the upper band has saturated.

This complementary behavior between the two emission bands extends the overall dynamic range of the sensor. The upper band provides high sensitivity at low concentrations, while the lower band maintains responsiveness at elevated concentrations where the upper band becomes saturated. Together, these bands produce a continuous and quantifiable sensing response across the full concentration range, a capability not achievable using a single-band emitter.

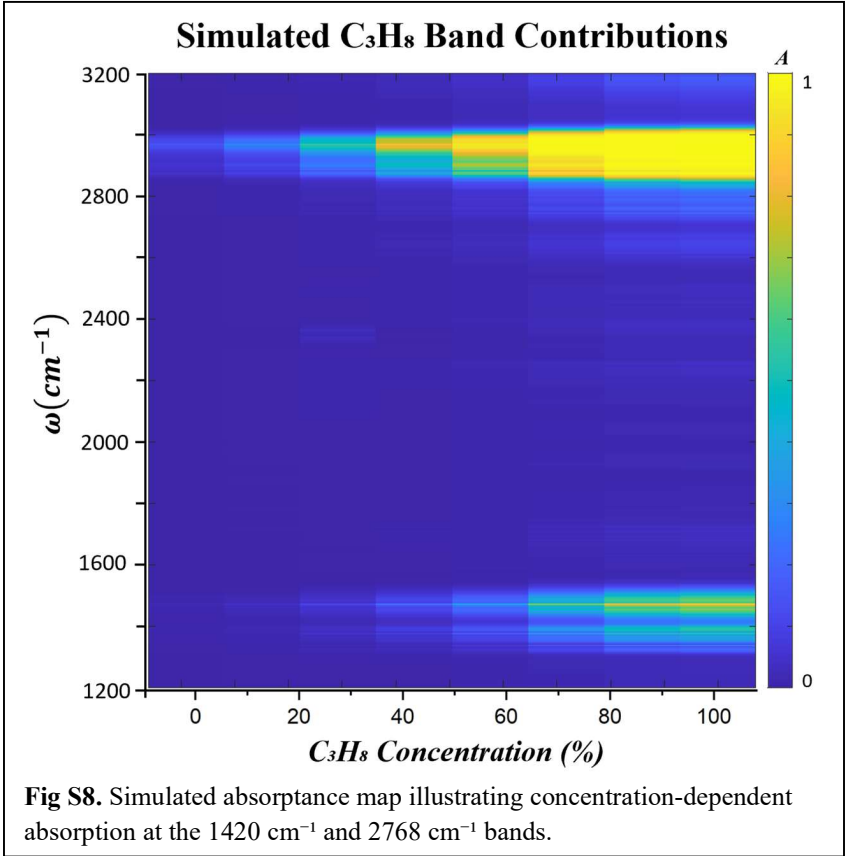

**Fig S8.** Simulated absorptance map illustrating concentration-dependent absorption at the 1420 cm$^{-1}$ and 2768 cm$^{-1}$ bands.

### S.7 k-Space Mapping of Dual-Band Emission

Momentum-resolved emission simulations were performed to visualize the spatial confinement and angular distribution of the two resonant modes supported by the dual-band C$_3$H$_8$ a-DBR emitter. **Figures S9(a - c)** show the normalized emission intensity as a function of the in-plane wavevectors ($k_x/k_0$ and $k_y/k_0$) for the lower band, upper band, and their combined response, respectively. The color scale ($\Omega$) represents the normalized emission strength across momentum space.

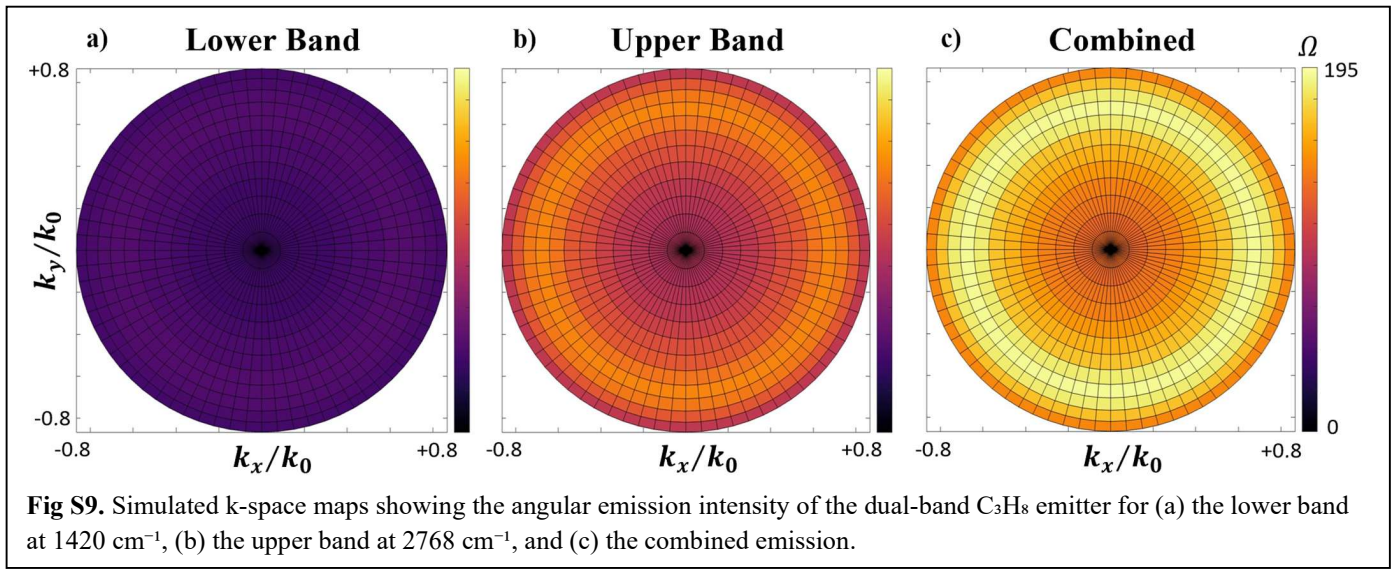

**Fig S9.** Simulated k-space maps showing the angular emission intensity of the dual-band C$_3$H$_8$ emitter for (a) the lower band at 1420 cm$^{-1}$, (b) the upper band at 2768 cm$^{-1}$, and (c) the combined emission.

Both the lower (1420 cm$^{-1}$) and upper (2768 cm$^{-1}$) bands exhibit strong confinement within the light line, confirming that the observed features correspond to surface-confined optical modes rather than freely propagating radiation. This behavior reflects the hybrid nature of the optical states, which arise from coupling between the photonic cavity and metallic interface.

The circular symmetry of the intensity distributions indicates that the emission is spatially isotropic in-plane, consistent with the planar geometry of the a-DBR structure. While previous figures characterized the spectral dependence of the emission, these momentum-space maps provide complementary spatial information, showing how the radiative intensity varies with in-plane momentum. The combined map (**Fig. S9c**) demonstrates that the dual-band a-DBR design enhances the overall optical density of states and broadens the accessible angular range of emission, both of which facilitate stronger coupling and improved detection efficiency.

## S.8 Temperature-Dependent Emission Stability

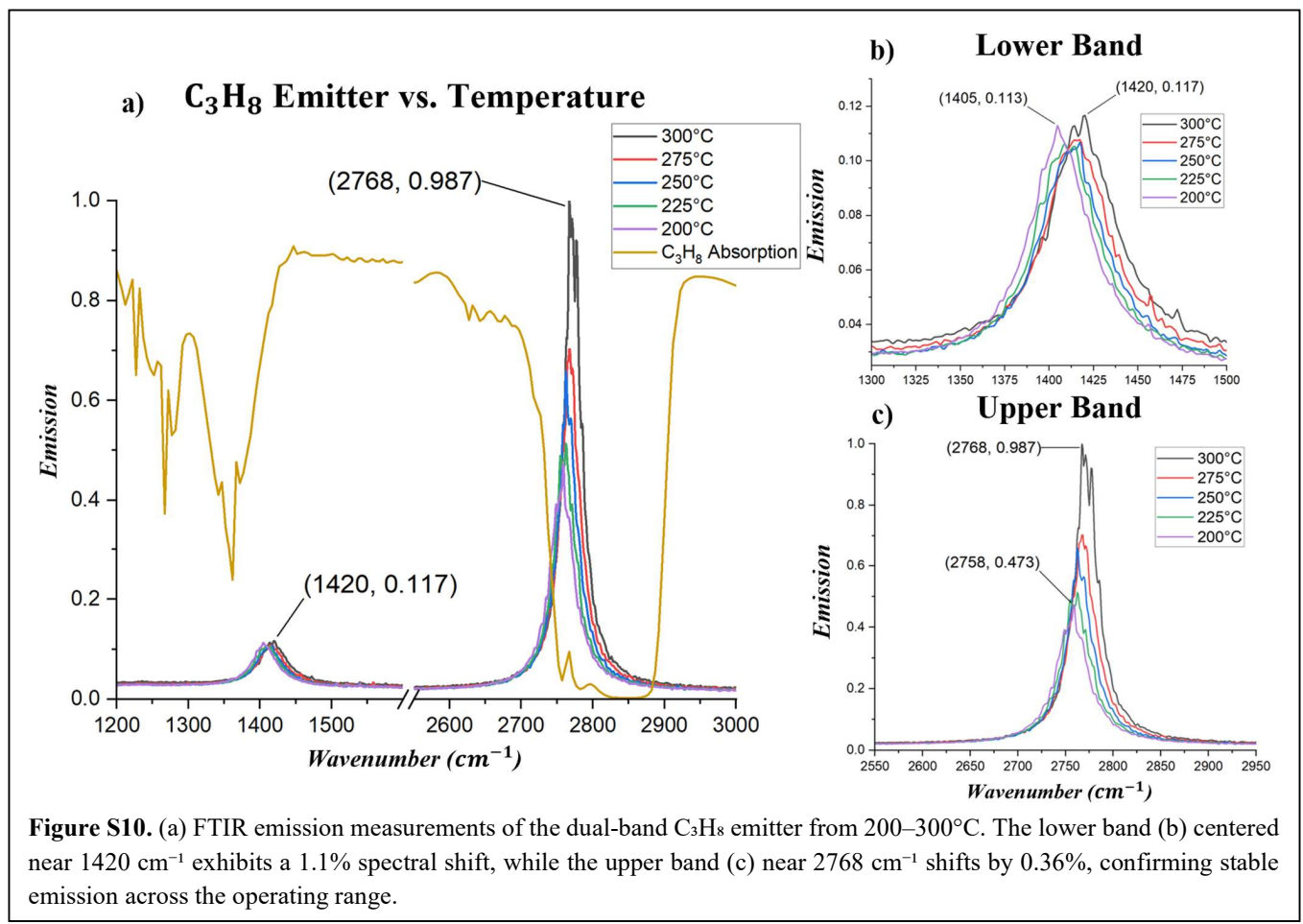

**Figure S10.** (a) FTIR emission measurements of the dual-band $C_3H_8$ emitter from 200–300°C. The lower band (b) centered near 1420 cm$^{-1}$ exhibits a 1.1% spectral shift, while the upper band (c) near 2768 cm$^{-1}$ shifts by 0.36%, confirming stable emission across the operating range.

Temperature-dependent emission spectra were collected for the CO, $CO_2$, and $C_3H_8$ emitters using Fourier-transform infrared (FTIR) spectroscopy and a Linkam heating stage. Emission measurements were performed from 200°C to 300°C in 25°C increments, allowing evaluation of thermal stability and resonance shifting with temperature.

Figures **S10(b)** and **S10(c)** show expanded views of the lower (1420 cm⁻¹) and upper (2768 cm⁻¹) emission bands of the dual-band $C_3H_8$ emitter presented in Figure **S10(a)**. Across all devices, the emission peaks exhibit only minor spectral shifts with increasing temperature, confirming excellent thermal stability of the emitters within the operational range. The magnitude of each shift was quantified using the relative change in peak frequency between 200°C and 300°C, normalized to the average resonance frequency:

$$\text{Percent shift} = \frac{|\tilde{v}_{300} - \tilde{v}_{200}|}{\tilde{v}_{avg}} \times 100$$

where $\tilde{v}_{300}$ and $\tilde{v}_{200}$ are the resonance frequencies at 300°C and 200°C, respectively.

For the $C_3H_8$ emitter, the lower-energy band shifted from 1405 cm⁻¹ at 200°C to 1420 cm⁻¹ at 300°C, corresponding to a 1.1% shift, while the upper-energy band shifted from 2758 cm⁻¹ to 2768 cm⁻¹, a 0.36% shift. Both resonances remain well aligned with the absorption bands of propane throughout the entire temperature range.

The CO emitter shifted from 2169 cm⁻¹ at 200°C to 2146 cm⁻¹ at 300°C, corresponding to a 1.1% shift, while the $CO_2$ emitter shifted from 2376 cm⁻¹ to 2351 cm⁻¹, also a 1.1% shift. These small shifts indicate that the emission frequencies remain spectrally confined within the respective gas absorption bands, ensuring consistent spectral overlap during sensing operation.

Overall, all emitters display less than 1.1% spectral variation across the 200°C – 300°C temperature range. This minimal shift confirms that the emitters are thermally stable and maintain spectral alignment with the target gas absorption bands under typical operating conditions.

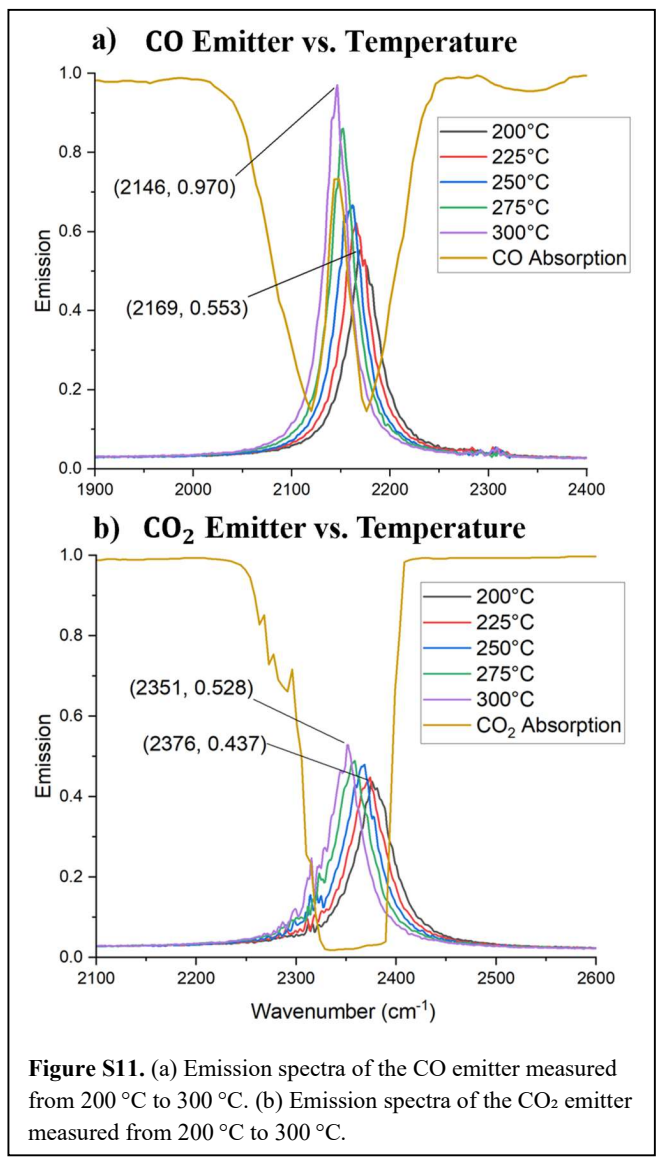

**Figure S11.** (a) Emission spectra of the CO emitter measured from 200 °C to 300 °C. (b) Emission spectra of the $CO_2$ emitter measured from 200 °C to 300 °C.

**S.8 Custom Tabletop Setup for Gas Sensing Measurements**

A custom optical setup was constructed to measure gas absorption using emission from the aperiodic distributed Bragg reflector (a-DBR) thermal emitters as seen in **Fig. S12.** The emitters were mounted on a Linkam THMS600 heating stage (Linkam Scientific Instruments, UK)[4] to enable controlled temperature operation. A custom gold-coated Winston cone (Optiforms, Inc.)[5] was positioned directly above the emitter to collimate the emitted radiation. The Winston cone featured an inner aperture of 8.0 mm, an entrance aperture (outer diameter) of 26.3 mm, and a parabolic reflective surface with a focal length of 7.9 mm. The a-DBR emitter was placed at the cone's focal point to ensure efficient collection and collimation of the emitted light.

A gold-coated mirror set at 45° incidence deflected the axial (z-direction) emission by 90°, redirecting it horizontally along the gas-cell optical axis. The collimated emission then entered a 10 cm gas cell equipped with $CaF_2$ windows (transmission range ≈ 50,000 $cm^{-1}$ – 800 $cm^{-1}$)[6]. The gas cell was connected to two Alicat mass-flow controllers that regulated the flow of calibration and analyte gases.[7] A 99.999% pure $N_2$ calibration gas and the target analyte gas were mixed in controlled ratios to achieve the desired concentration levels.

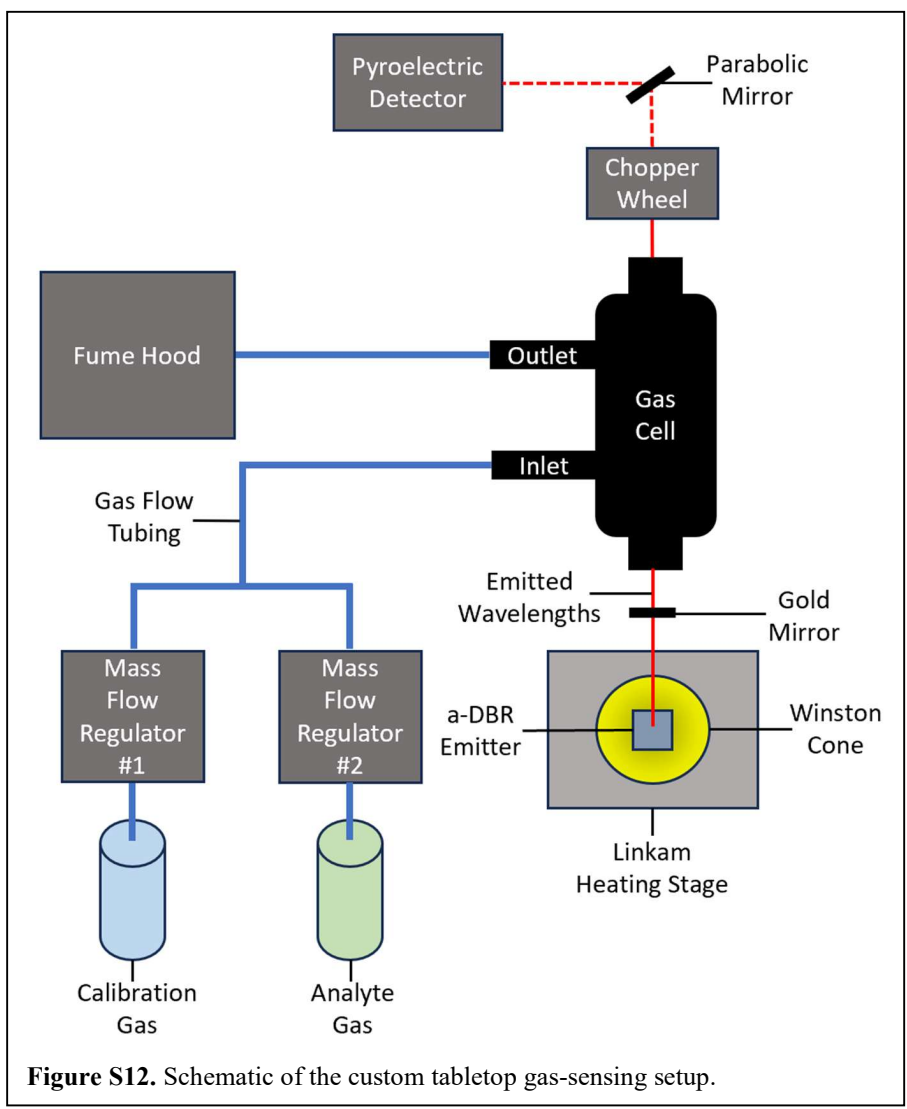

**Figure S12.** Schematic of the custom tabletop gas-sensing setup.

The gas cell was initially purged with $N_2$ for 3 minutes prior to measurement. The flow ratio was then adjusted in 5% increments, beginning at 0 sccm analyte/200 sccm $N_2$ and increasing to 200 sccm analyte/0 sccm $N_2$. Each new concentration was allowed to flow for 30 seconds before recording a 60-second emission measurement. All gases were delivered through Tygon gas-transfer tubing secured with Prestolok fittings, and exhaust gases were safely vented into a fume hood.

After transmission through the gas cell, the emission passed through a mechanical chopper wheel, which modulated the continuous signal prior to detection. The modulated beam was then focused by a 1-inch diameter, 90° off-axis parabolic mirror with a protected silver coating and a reflected focal length (RFL) of 6 inches[8], directing the light onto a pyroelectric detector[9] positioned at the mirror's focal distance. The pyroelectric detector output (in watts) corresponded to the transmitted emission intensity and was recorded at a sampling rate of approximately 15 measurements per second, enabling quantification of gas concentration via Beer's law.

## S.9 Beer Lambert Law

The Beer–Lambert law describes the exponential attenuation of light as it passes through an absorbing medium, relating the transmitted intensity to the concentration and optical path length of the absorber. It is expressed as:

$$I = I_0 e^{-\alpha c L}$$

where $I_0$ and $I$ are the incident and transmitted intensities, respectively, $\alpha$ is the absorption coefficient of the analyte, $c$ is the concentration, and $L$ is the optical path length. In logarithmic form, the absorbance $A$ is given by:

$$A = \log_{10}\left(\frac{I_0}{I}\right) = \varepsilon c L$$

where $\varepsilon$ is the molar absorptivity, a material- and wavelength-dependent constant.

In this work, the Beer–Lambert law was applied to quantify gas concentration from the measured emission intensity transmitted through the 10 cm gas cell. The pyroelectric detector measured the transmitted power ($I$) at each analyte concentration, while the initial power in pure $N_2$ ($I_0$) served as the reference intensity. The resulting absorbance spectra were then computed from the ratio $I_0/I$, allowing the relative change in transmitted intensity to be directly correlated to the analyte concentration. This relationship enabled quantitative assessment of gas absorption for CO, $CO_2$, and $C_3H_8$ using the a-DBR emitter as the tunable radiation source.

## S.10 Power calculations

A semi-analytical framework was developed to estimate the concentration-dependent transmitted power and minimum detection limit of the gas-sensing setup. The calculation incorporates the molar absorptivity of the target gas, the temperature- and angle-dependent emissivity of the emitter, the collection efficiency of the optical system, and the responsivity and noise characteristics of the detector.

### S.10.1 Molar Absorptivity

The molar absorptivity ($\varepsilon$) of each gas species was derived from transmission spectra using the Beer–Lambert law:

$$-T = \varepsilon L c$$

where $T$ is the measured transmission, $L$ is the gas-cell path length (10 cm), and $c$ is the gas concentration. From these measurements, $\varepsilon$ was determined and used to estimate absorbance as a function of concentration and optical path length.

### S.10.2 Emitter Radiated Power

The spectral power radiated by the a-DBR emitter was modeled as:

$$P_r = \int_0^{2\pi} \int_0^{\phi_{max}} \int_{\lambda_0}^{\lambda_{max}} I_{BB}(T, \lambda)\, \varepsilon(\lambda, \theta) \sin\theta \cos\theta\, d\lambda\, d\theta\, d\phi$$

where $I_{BB}(T, \lambda)$ is the Planck blackbody intensity, and $\varepsilon(\lambda, \theta)$ is the angle-dependent emissivity obtained experimentally via FTIR emission measurements. The limits $\phi_{max}$ and $\lambda_{max}$ are determined by the geometry and optical bandwidth of the Winston cone and gas cell, respectively.

For an ideal blackbody,

$$I_{BB}(T, \lambda) = \frac{2hc^2}{\lambda^5}\left(e^{\frac{hc}{\lambda k_B T}} - 1\right)^{-1}$$

and the total radiated power is modulated by the emitter's emissivity spectrum $\varepsilon(\lambda, \theta)$.

### S.10.3 Optical Transfer Function

Optical throughput was modeled via a multiplicative transfer function representing reflections and transmissions through all optical components:

$$\tau(\lambda, c) = C\, R_{m1}\, T_{w1}(\lambda)\, T_{gc}(\lambda, c)\, T_{w2}(\lambda)\, R_{bs}\, R_{p1}$$

where $R_m$ and $T_w$ denote reflection and transmission coefficients of the mirrors and windows, respectively, and $T_{gc}(\lambda, c)$ represents gas-cell transmission at concentration $c$. The calibration constant $C$ accounts for unmodeled DC offsets and normalization to experimental data. The total detected power is then expressed as:

$$P_d(c) = \int_0^{2\pi} \int_0^{\phi_{max}} \int_{\lambda_0}^{\lambda_{max}} I_{BB}(T, \lambda)\, \varepsilon(\lambda, \theta)\, \tau(\lambda, c) \sin\theta \cos\theta\, d\lambda\, d\theta\, d\phi$$

### S.10.4 Detector Noise and Sensitivity

The pyroelectric detector used in this work (Ophir Optics RM9, with lock-in amplifier)[9] has a noise-equivalent power (NEP) of approximately 95 nW·Hz$^{-1/2}$, derived from the manufacturer's specified power noise level of 30 nW averaged over a 10 s interval. The effective detection range of the system is 100 nW to 100 mW, with an 8 mm aperture diameter. The minimum detectable signal ($P_{\min}$) was estimated by equating the concentration-dependent differential power change ($\Delta P_d$) to the detector NEP, providing an estimate of the minimum detectable gas concentration under the experimental conditions.

## S.11 Comparison to Commercial NDIR Gas Sensor

This section evaluates the enhancement in sensitivity achieved by the dual-band $C_3H_8$ aperiodic distributed Bragg reflector (a-DBR) emitter compared to a commercial single-band narrowband infrared (NDIR) gas sensor. The a-DBR emitter was specifically designed to provide simultaneous emission at two spectrally tailored resonances that align with the mid-infrared absorption features of propane ($C_3H_8$) near 2768 cm$^{-1}$ and 1420 cm$^{-1}$, enabling direct comparison of dual- versus single-band performance.

The commercial InfraTec NDIR filter is centered at 2679 cm$^{-1}$ (3.73 μm) with a full width at half maximum (FWHM) of 63.5 cm$^{-1}$ and a peak transmission of 0.80.[10] In contrast, the custom a-DBR emitter exhibits dual-band emission centered at 2768 cm$^{-1}$ and 1420 cm$^{-1}$, with FWHM values of 29.5 cm$^{-1}$ and 61.2 cm$^{-1}$, respectively, and corresponding peak emissivities of 0.987 and 0.117.

To quantify the relative sensitivity, the detected power was modeled as the source spectrum weighted by the gas transmission,

$$T(\tilde{v}; C, L) = \exp[-\kappa(\tilde{v})\, C\, L] \approx 1 - \kappa(\tilde{v})\, C\, L,$$

where $\kappa(\tilde{v})$ is the propane absorption coefficient, $C$ is concentration, and $L$ is the gas-cell path length (10 cm). The fractional power change was then expressed as

$$\frac{\Delta P}{P_0} \approx \int W(\tilde{v})\, \kappa(\tilde{v})\, d\tilde{v}\, C\, L = S\, C\, L,$$

where $W(\tilde{v})$ represents either the emitter emissivity or the InfraTec filter transmission spectrum. The proportionality constant $S$ defines the spectral sensitivity, and the enhancement ratio was calculated as:

$$E = \frac{S_{\text{a-DBR}}}{S_{\text{InfraTec}}}.$$

Two cases were evaluated to examine the impact of the emitter's enhanced emissivity and its dual-band emission capability. The single-band enhancement compares emission near 2768 cm$^{-1}$ to the InfraTec filter response, isolating improvements arising from the stronger and more spectrally matched emissivity of the a-DBR design. The dual-band enhancement, on the other hand, demonstrates the core advancement of this work, the ability to emit at two distinct resonance frequencies simultaneously, enabling multi-band detection within a single emitter and expanding sensing capability beyond what is possible with traditional single-band NDIR filters.

Using the InfraTec filter parameters (center = 2679 cm$^{-1}$, FWHM = 63.5 cm$^{-1}$), the integrated sensitivities were:

$$S_{\text{InfraTec}} = 8.7588, \quad S_{2768} = 8.0967, \quad S_{\text{dual}} = 11.9565$$

yielding

$$E_{\text{single}} = 0.92, \quad E_{\text{dual}} = 1.36$$

In this configuration, the commercial filter encompasses a broader spectral region, capturing a larger portion of the propane absorption band and therefore exhibiting inherently higher apparent sensitivity. However, this increased bandwidth also reduces chemical selectivity by broadening the detection window, increasing the likelihood of false positives from nearby gas species that are spectrally close.

To ensure a fair and physically consistent comparison, the InfraTec filter was recalculated using a matched FWHM of 29.5 cm$^{-1}$, equivalent to the linewidth of the a-DBR emitter's 2768 cm$^{-1}$ resonance. This bandwidth matching normalizes the comparison so that the calculated enhancement reflects differences in emitter design rather than filter width. Under these matched conditions,

$$S_{\text{InfraTec, 29.5}} = 4.2090, \quad S_{2768} = 8.0967, \quad S_{\text{dual}} = 11.9565$$

producing enhancement ratios of

$$E_{\text{single, matched}} = 1.92, \quad E_{\text{dual, matched}} = 2.84$$

These results confirm that the single-band $C_3H_8$ a-DBR emitter already provides nearly a 1.92× improvement in sensitivity due to its higher emissivity, while the dual-band configuration further increases this enhancement to 2.84× by enabling simultaneous emission at multiple resonances within a single device. Together, these findings demonstrate that the a-DBR architecture enables a new class of multi-band emitters that improve sensitivity without compromising spectral selectivity, achieving performance beyond that of conventional NDIR systems.

# References

(1) Joannopoulos, J. D.; Johnson, S. G.; Winn, J. N.; Meade, R. D. *Photonic Crystals: Molding the Flow of Light - Second Edition*; Princeton University Press, 2011. https://doi.org/10.1515/9781400828241.

(2) He, M.; Nolen, J. R.; Nordlander, J.; Cleri, A.; McIlwaine, N. S.; Tang, Y.; Lu, G.; Folland, T. G.; Landman, B. A.; Maria, J.-P.; Caldwell, J. D. Deterministic Inverse Design of Tamm Plasmon Thermal Emitters with Multi-Resonant Control. *Nat. Mater.* **2021**, *20* (12), 1663–1669. https://doi.org/10.1038/s41563-021-01094-0.

(3) mingze321. Mingze321/Tamm, 2021. https://github.com/mingze321/Tamm (accessed 2025-10-23).

(4) *Linkam THMS600 - Temperature Control Stage for Microscopy and Spectroscopy*. Linkam Scientific. https://www.linkam.co.uk/thms600 (accessed 2025-10-23).

(5) *Electroformed Parabolic Reflectors*. https://www.optiforms.com/electroformed-components/parabolic-reflectors/ (accessed 2025-10-23).

(6) PIKE-Technologies_Stainless-Steel-Short-Path-Gas-Cells.Pdf. https://www.piketech.com/wp-content/uploads/PDS/transmission/PIKE-Technologies_Stainless-Steel-Short-Path-Gas-Cells.pdf (accessed 2025-10-23).

(7) *Alicat Mass Flow Controller with Display, 0 - 200 SCCM from Cole-Parmer*. https://www.coleparmer.com/i/alicat-mass-flow-controller-with-display-0-200-sccm/1530902?PubID=UX&persist=true&ip=no&gad_source=1&gad_campaignid=21168216896&gbraid=0AAAAAD-n1iXfA7D34vlhqAd6H1yLgJlXn&gclid=CjwKCAjwpOfHBhAxEiwAm1SwEhJG4K7x0MfHt-KoKNBrwos-GO67HQvQCzqhLueQXDDFuk4gWhg0nBoCAdAQAvD_BwE (accessed 2025-10-23).

(8) *Thorlabs - MPD169-P01 Ø1*. https://www.thorlabs.com (accessed 2025-10-23).

(9) *RM9-Pyro 100 nW to 100 mW Pyroelectric Sensor Radiometer*. https://www.ophiropt.com/en/f/rm9-pyro-radiometer (accessed 2025-10-23).

(10) Standardfilter_Katalog+homepage_Layout.Xlsx. https://media.infratec.eu/infratec-b-filter-homepage-layout.pdf?mp_enc=bXBfZGlyPTY1MTY3Jm1wX2ZpbGU9NDc3OTM0MjM= (accessed 2025-10-23).